# Lasserre Hierarchy, Higher Eigenvalues, and Approximation Schemes for Quadratic Integer Programming with PSD Objectives

Venkatesan Guruswami[*]     Ali Kemal Sinop[*]

Computer Science Department
Carnegie Mellon University
Pittsburgh, PA 15213.

**Abstract**

We present an approximation scheme for optimizing certain Quadratic Integer Programming problems with positive semidefinite objective functions and global linear constraints. This framework includes well known graph problems such as Minimum graph bisection, Edge expansion, Uniform sparsest cut, and Small Set expansion, as well as the Unique Games problem. These problems are notorious for the existence of huge gaps between the known algorithmic results and NP-hardness results. Our algorithm is based on rounding semidefinite programs from the Lasserre hierarchy, and the analysis uses bounds for low-rank approximations of a matrix in Frobenius norm using columns of the matrix.

For all the above graph problems, we give an algorithm running in time $n^{O(r/\varepsilon^2)}$ with approximation ratio $\frac{1+\varepsilon}{\min\{1,\lambda_r\}}$, where $\lambda_r$ is the $r$'th smallest eigenvalue of the normalized graph Laplacian $\mathcal{L}$. In the case of graph bisection and small set expansion, the number of vertices in the cut is within lower-order terms of the stipulated bound. Our results imply $(1 + O(\varepsilon))$ factor approximation in time $n^{O(r^*/\varepsilon^2)}$ where $r^*$ is the number of eigenvalues of $\mathcal{L}$ smaller than $1 - \varepsilon$. This perhaps gives some indication as to why even showing mere APX-hardness for these problems has been elusive, since the reduction must produce graphs with a slowly growing spectrum (and classes like planar graphs which are known to have such a spectral property often admit good algorithms owing to their nice structure).

For Unique Games, we give a factor $(1 + \frac{2+\varepsilon}{\lambda_r})$ approximation for minimizing the number of unsatisfied constraints in $n^{O(r/\varepsilon)}$ time. This improves an earlier bound for solving Unique Games on expanders, and also shows that Lasserre SDPs are powerful enough to solve well-known integrality gap instances for the basic SDP.

We also give an algorithm for independent sets in graphs that performs well when the Laplacian does not have too many eigenvalues bigger than $1 + o(1)$.

---

[*]Research supported in part by a Packard Fellowship, NSF CCF 0963975, and US-Israel BSF grant 2008293. Email: guruswami@cmu.edu, asinop@cs.cmu.edu

# Contents

# 1 Introduction

The theory of approximation algorithms has made major strides in the last two decades, pinning down, for many basic optimization problems, the exact (or asymptotic) threshold up to which efficient approximation is possible. Some notorious problems, however, have withstood this wave of progress; for these problems the best known algorithms deliver super-constant approximation ratios, whereas NP-hardness results do not even rule out say a factor $1.1$ (or sometimes even a factor $(1+\varepsilon)$ for any constant $\varepsilon > 0$) approximation algorithm. Examples of such problems include graph partitioning problems such as minimum bisection, uniform sparsest cut, and small-set expansion; finding a dense subgraph induced on $k$ vertices; minimum linear arrangement; and constraint satisfaction problems such as minimum CNF deletion or Unique Games.

There has been evidence of three distinct flavors for the hardness of these problems: (i) Ruling out a polynomial time approximation scheme (PTAS) assuming that $\mathsf{NP} \not\subset \bigcap_{\varepsilon>0} \mathsf{BPTIME}(2^{n^{\varepsilon}})$ via quasi-random PCPs [Kho06, AMS07]; (ii) Inapproximability results within some constant factor assuming average-case hardness of refuting random 3SAT instances [Fei02]; and (iii) Inapproximability within super-constant factors under a strong conjecture on the intractability of the small-set expansion (SSE) problem [RST10]. While (iii) gives the strongest hardness results, it is conditioned on the conjectured hardness of SSE [RS10], an assumption that implies the Unique Games conjecture, and arguably does not yet have as much evidence in its support as the complexity assumptions made in (i) or (ii).

In this work, we give a unified algorithm, based on powerful semidefinite programs from the Lasserre hierarchy, for several of these problems, and a broader class of quadratic integer programming problems with linear constraints (more details are in Section 1.1 below). Our algorithms deliver a good approximation ratio if the eigenvalues of the Laplacian of the underlying graph increase at a reasonable rate. In particular, for all the above graph partitioning problems, we get a $(1+\varepsilon)/\min\{\lambda_r, 1\}$ approximation factor in $n^{O_\varepsilon(r)}$ time, where $\lambda_r$ is the $r$'th smallest eigenvalue of the normalized Laplacian (which has eigenvalues in the interval $[0, 2]$). Note that if $\lambda_r \geqslant 1 - \varepsilon$, then we get a $(1 + O(\varepsilon))$ approximation ratio.

**Perspective.** The direct algorithmic interpretation of our results is simply that one can probably get good approximations for graphs that are pretty "weak-expanders," in that we only require lower bounds on higher eigenvalues rather than on $\lambda_2$ as in the case of expanders. In terms of our broader understanding of the complexity of approximating these problems, our results perhaps point to why even showing APX-hardness for these problems has been difficult, as the reduction must produce graphs with a very slowly growing spectrum, with many ($n^{\Omega(1)}$, or even $n^{1-o(1)}$ for near-linear time reductions) small eigenvalues. Trivial examples of such graphs are the disjoint union of many small components (taking the union of $r$ components ensures $\lambda_r = 0$), but these are of course easily handled by working on each component separately. We note that Laplacians of planar graphs, bounded genus graphs, and graphs excluding fixed minors, have many small eigenvalues [KLPT10], but these classes are often easier to handle algorithmically due to their rich structure — for example, conductance and edge expansion problems are polynomial time solvable on planar graphs [PP93]. Also, the recent result of [ABS10] shows that if $\lambda_r = o(1)$ for some $r = n^{\Omega(1)}$, then the graph must have an $n^{1-\Omega(1)}$ sized subset with very few edges leaving it. Speculating somewhat boldly, may be these results suggest that graphs with too many small eigenvalues are also typically not hard instances for these problems.

Our results also give some explanation for our inability so far to show integrality gaps for even 4 rounds of the Lasserre hierarchy for problems which we only know to be hard assuming the Unique Games conjecture (UGC). In fact, it is entirely consistent with current knowledge that just $O(1)$ rounds of the Lasserre hierarchy gives an improvement over the $0.878$ performance ratio of the Goemans-Williamson algorithm for Max Cut, and refutes the UGC!

## 1.1 Summary of results

Let us now state our specific results informally.

**Graph partitioning.** We begin with results for certain cut/graph partitioning problems. For simplicity, we state the results for unweighted graphs — the body of the paper handles weighted graphs. Below $\lambda_p$ denotes the $p$'th smallest eigenvalue of the normalized Laplacian $\mathcal{L}$ of the graph $G$, defined as $\mathcal{L} = D^{-1/2}(I - A)D^{-1/2}$ where $A$ is the adjacency matrix and $D$ is a diagonal matrix with node degrees on the diagonal. (In the stated approximation ratios, $\lambda_r$ (resp. $2 - \lambda_{n-r}$) should be understood as $\min\{\lambda_r, 1\}$ (resp. $\min\{2 - \lambda_{n-r}, 1\}$), but we don't make this explicit to avoid notational clutter.) The algorithm's running time is $n^{O(r/\varepsilon^2)}$ in each case. This runtime arises due to solving the standard semidefinite programs (SDP) lifted with $O(r/\varepsilon^2)$ rounds of the Lasserre hierarchy. Our results are shown via an efficient rounding algorithm whose runtime is $n^{O(1)}$; the exponential dependence on $r$ is thus limited to solving the SDP.

- MAXIMUM CUT AND MINIMUM UNCUT: Given a graph $G$ on $n$ vertices with a partition leaving at most $b$ many edges uncut, we can find a partition that leaves at most $\frac{1+\varepsilon}{2-\lambda_{n-r}}b$ many edges uncut. (We can also get an approximation guarantee of $(1 + \frac{2 \pm \varepsilon}{\lambda_r})$ for Minimum Uncut as a special case of our result for Unique Games.)

- MINIMUM (MAXIMUM) BISECTION: Given a graph $G$ on $n$ vertices with a bisection (partition into two equal parts) cutting (uncutting) at most $b$ edges, we can find a near-bisection, with each side having $\frac{n}{2} \pm O_\varepsilon(\sqrt{n})$ vertices, that cuts at most $\frac{1+\varepsilon}{\lambda_r}b$ (uncuts at most $\frac{1+\varepsilon}{2-\lambda_{n-r}}b$) edges respectively.

- SMALL-SET EXPANSION (SSE): Given a graph $G$ on $n$ vertices with a set of volume $\mu$ with at most $b$ edges leaving it, we can find a set $U$ of volume $\mu \pm O_\varepsilon(\sqrt{d_{\max}\mu})$ with at most $\frac{1+\varepsilon}{\lambda_r}b$ edges leaving $U$. We can also find such a set with volume $\mu(1 \pm \varepsilon)$, which will be a better guarantee for highly irregular graphs.

- Various graph partitioning problems which involve minimizing ratio of cut size with size or volume of partitions. For each problem below, we can find a non-empty set $U \subsetneq V$, whose value is at most $\frac{1+\varepsilon}{\lambda_r}$OPT.

    - UNIFORM SPARSEST CUT: $\phi_G(U)$ — defined as the ratio of the number of edges in the cut $(U, V \setminus U)$ divided by $|U||V \setminus U|$.
    - EDGE EXPANSION: $h_G(U)$ — defined as the ratio of the number of edges leaving $U$ to the number of nodes in $U$, where $U$ is the smaller side of the cut.
    - NORMALIZED CUT: $\text{ncut}_G(U)$ — defined as the ratio of the number of edges in the cut $(U, V \setminus U)$ divided by the product of the volumes of $U$ and $V \setminus U$.

– CONDUCTANCE: conductance$_G(U)$ — defined as the fraction of edges incident on $U$ that leave $U$ where $U$ is the side of the cut with smaller volume.

In each case, we can also handle boundary conditions stipulating that $U$ must contain some subset $F$ of nodes, and avoid some other disjoint subset $B$ of nodes. This feature is used for our results on the above ratio-minimization objectives as well as finding a set of volume $\mu(1 \pm \varepsilon)$ for SSE.

We can give similar guarantees for the $k$-way partitioning versions of these problems (when the version makes sense). For example, for the $k$-way section problem of splitting the vertices into $k$ equal parts to minimize the number of cut edges, we can give a $(1 + \varepsilon)/\lambda_r$ approximation in $n^{O(kr/\varepsilon^2)}$ time, with $O(\sqrt{n \log(k/\varepsilon)})$ error in the size of each part.

**Remark 1** (Subspace enumeration). We note that for conductance (and related problems with quotient objectives mentioned above), it is possible to get a $O(1/\lambda_r)$ approximation in $n^{O(r)}$ time by searching for a good cut in the $r$-dimensional eigenspace corresponding to the $r$ smallest eigenvalues (we thank David Steurer for pointing out that this is implicit in the subspace enumeration results of [ABS10]). It is not clear, however, if such methods can give a $(1+\varepsilon)/\lambda_r$ type ratio. Further, this method does not apply in the presence of a balance requirement such as Minimum Bisection, as it could violate the balance condition by an $\Omega(n)$ amount, or for Unique Games (where we do not know how to control the spectrum of the lifted graph). □

**PSD Quadratic Integer Programs.** In addition to the above cut problems, our method applies more abstractly to the class of minimization quadratic integer programs (QIP) with positive semidefinite (PSD) cost functions and arbitrary linear constraints.

- QIP WITH PSD COSTS: Given a PSD matrix $A \in \mathbb{R}^{(V \times [k]) \times (V \times [k])}$, consider the problem of finding $\widetilde{x} \in \{0,1\}^{V \times [k]}$ minimizing $\widetilde{x}^T A \widetilde{x}$ subject to: (i) exactly one of $\{\widetilde{x}_u(i)\}_{i \in [k]}$ equals 1 for each $u$, and (ii) the linear constraints $B\widetilde{x} \geq c$. We find such an $\widetilde{x}$ with $\widetilde{x}^T L \widetilde{x} \leq \frac{1+\varepsilon}{\min\{1, \lambda_r(\mathcal{A})\}}$ where $\mathcal{A} = \mathrm{diag}(A)^{-1/2} \cdot A \cdot \mathrm{diag}(A)^{-1/2}$.

**Unique Games.** We next state our result for Unique Games. This is not a direct application of the result for QIP; see Section 1.2 for details on the difficulties.

- UNIQUE GAMES: Given a Unique Games instance with constraint graph $G = (V, E)$, label set $[k]$, and bijective constraints $\pi_e$ for each edge, if the optimum assignment $\sigma : V \to [k]$ fails to satisfy $\eta$ of the constraints, we can find an assignment that fails to satisfy at most $\eta \left(1 + \frac{2+\varepsilon}{\lambda_r}\right)$ of the constraints.

In this case, we are only able to get a weaker $\approx 1 + 2/\lambda_r$ approximation factor, which is always larger than 2. In this context, it is interesting to note that minimizing the number of unsatisfied constraints in Unique Games *is* known to be APX-hard; for example, the known NP-hardness for approximating Max Cut [Hås01, TSSW00] implies a factor $(5/4-\varepsilon)$ hardness for this problem (and indeed for the special case of Minimum Uncut).

**Remark 2** (UG on expanders). Arora *et al* [AKK$^+$08] showed that Unique Games is easy on expanders, and gave an $O(\frac{\log(1/\mathtt{OPT})}{\lambda_2})$ approximation to the problem of minimizing the number of unsatisfied constraints, where OPT is the fraction of unsatisfied constraints in the optimal solution. For the subclass of "linear" Unique Games, they achieved an approximation ratio of $O(1/\lambda_2)$ without any dependence on OPT. A factor $O(1/\lambda_2)$ approximation ratio was achieved for general Unique Games instances by Makarychev and Makarychev [MM10] (assuming $\lambda_2$ is large enough, they also get a $O(1/h_G)$ approximation where $h_G$ is the Cheeger constant). Our result achieves an approximation factor of $O(1/\lambda_r)$, if one is allowed $n^{O(r)}$ time.

For instances of ΓMAX2LIN, the paper [AKK$^+$08] also gives an $n^{O(r)}$ time algorithm that satisfies all but a fraction $O(\text{OPT}/z_r(G))$ of constraints, where $z_r(G)$ is the value of the $r$-round Lasserre SDP relaxation of Sparsest Cut on $G$. For $r = 1$, $z_1(G) = \lambda_2$. But the growth rate of $z_r(G)$, eg. its relation to the Laplacian spectrum, was not known. □

**Remark 3** (SDP gap instances). Our algorithm also shows that the Khot-Vishnoi UG gap instance for the basic SDP [KV05], has $O(1)$ integrality gap for the lifted SDP corresponding to $\text{poly}(\log n)$ rounds of Lasserre hierarchy. In particular, these instances admit quasi-polynomial time constant factor approximations. This latter result is already known and was shown by Kolla [Kol10] using spectral techniques. Our result shows that strong enough SDPs also suffice to tackle these instances. In a similar vein, applying the ABS graph decomposition [ABS10] to split the graph into components with at most $n^\varepsilon$ small eigenvalues while cutting very few edges, one also gets that $n^{\varepsilon^{\Omega(1)}}$ rounds of the Lasserre hierarchy suffice to well-approximate Unique Games on instances with at most $\varepsilon$ fraction unsatisfied constraints. □

**Independent Set in graphs.** We also give a rounding algorithm for the natural Lasserre SDP for independent set in graphs. Here, our result gives an algorithm running in $n^{O(r/\varepsilon^2)}$ time algorithm that finds an independent set of size $\approx \frac{n}{2d_{\max}} \frac{1}{\lambda_{n-r}-1}$ where $\lambda_{n-r} > 1$ is the $r$'th largest eigenvalue of the graph's normalized Laplacian. Thus even exact independent set is easy for graphs for which the number of eigenvalues greater than $\approx 1 + \frac{1}{4d_{\max}}$ is small.

## 1.2 Our Techniques

Our results follow a unified approach, based on a SDP relaxation of the underlying integer program. The SDP is chosen from the Lasserre hierarchy [Las02], and its solution has vectors $x_T(\sigma)$ corresponding to local assignments to every subset $T \subset V$ of at most $r'$ vertices. (Such an SDP is said to belong to $r'$ rounds of the Lasserre hierarchy.) The vectors satisfy dot product constraints corresponding to consistency of pairs of these local assignments. (See Section 2 for a formal description.)

Given an optimal solution to the Lasserre SDP, we give a rounding method based on local propagation, similar to the rounding algorithm for Unique Games on expanders in [AKK$^+$08]. We first find an appropriate subset $S$ of $r'$ nodes (called the *seed* nodes). One could simply try all such subsets in $n^{r'}$ time, though there is an $O(n^5)$ time algorithm to locate the set $S$ as well. Then for each assignment $f$ to nodes in $S$, we randomly extend the assignment to all nodes by assigning, for each $u \in V \setminus S$ independently, a random value from $u$'s marginal distribution based on $x_{S \cup \{u\}}$ conditioned on the assignment $f$ to $S$.

After arithmetizing the performance of the rounding algorithm, and making a simple but crucial observation that lets us pass from higher order Lasserre vectors to vectors corresponding to single vertices, the core step in the analysis is the following: Given vectors $\{X_v \in \mathbb{R}^\Upsilon\}_{v \in V}$ and an upper bound on a positive semidefinite (PSD) quadratic form $\sum_{u,v \in V} L_{uv} \langle X_u, X_v \rangle = \mathsf{Tr}(X^T X L) \leqslant \eta$, place an upper bound on the sum of the squared distance of $X_u$ from the span of $\{X_s\}_{s \in S}$, i.e., the quantity $\sum_u \|X_S^\perp X_u\|^2 = \mathsf{Tr}(X^T X_S^\perp X)$. (Here $X \in \mathbb{R}^{\Upsilon \times V}$ is the matrix with columns $\{X_v : v \in V\}$.)

We relate the above question to the problem of *column-selection* for low-rank approximations to a matrix, studied in many recent works [DV06, DR10, BDMI11, GS11]. It is known by the recent works [BDMI11, GS11][1] that one can pick $r/\varepsilon$ columns $S$ such that $\mathsf{Tr}(X^T X_S^\perp X)$ is at most $1/(1-\varepsilon)$ times the error of the best rank-$r$ approximation to $X$ in Frobenius norm, which equals $\sum_{i>r} \sigma_i$ where the $\sigma_i$'s are the eigenvalues of $X^T X$ in decreasing order. Combining this with the upper bound $\mathsf{Tr}(X^T X L) \leqslant \eta$, we deduce an approximation ratio of $\left(1 + \frac{1+\varepsilon}{\lambda_{r+1}}\right)$ for our algorithm. Also, the independent rounding of each $u$ implies, by standard Chernoff-bounds, that any linear constraint (such as a balance condition) is met up to lower order deviations.

Note that the above gives an approximation ratio $\approx 1 + 1/\lambda_r$, which always exceeds 2. To get our improved $(1+\varepsilon)/\lambda_r$ guarantee, we need a more refined analysis, based on iterated application of column selection along with some other ideas.

For Unique Games, a direct application of our framework for quadratic IPs would require relating the spectrum of the constraint graph $G$ of the Unique Games instance to that of the lifted graph $\widehat{G}$. There are such results known for random lifts, for instance [Oli10]; saying something in the case of arbitrary lifts, however, seems very difficult.[2] We therefore resort to an indirect approach, based on *embedding* the set of $k$ vectors $\{x_u(i)\}_{i \in [k]}$ for a vertex into a single vector $X_u$ with some nice distance preserving properties that enables us to relate quadratic forms on the lifted graph to a proxy form on the base constraint graph. This idea was also used in [AKK+08] for the analysis of their algorithm on expanders, where they used an embedding based on non-linear tensoring. In our case, we need the embedding to also preserve distances from certain higher-dimensional subspaces (in addition to preserving pairwise distances); this favors an embedding that is as "linear" as possible, which we obtain by passing to a tensor product space.

### 1.3 Related work on Lasserre SDPs in approximation

The Lasserre SDPs seem very powerful, and as mentioned earlier, for problems shown to be hard assuming the UGC (such as beating Goemans-Williamson for Max Cut), integrality gaps are not known even for a small constant number of rounds. A gap instance for Unique Games is known if the Lasserre constraints are only *approximately* satisfied [KPS10]. It is interesting to contrast this with our positive result. The error needed in the constraints for the construction in [KPS10] is $r/(\log \log n)^c$ for some $c < 1$, where $n$ is the number of vertices and $r$ the number of rounds. Our analysis requires the Lasserre consistency constraints are met exactly. In Appendix A, we present an algorithm that produces such valid Lasserre SDP solutions in time $(kn)^{O(r)} O(\log(1/\varepsilon_0))$ with an additive error of $\varepsilon_0$ in linear constraints, and an objective value at most $\varepsilon_0$ more than optimal.

Strong Lasserre integrality gaps have been constructed for certain approximation problems that *are* known to be NP-hard. Schoenebeck proved a strong negative result that even $\Omega(n)$ rounds of the Lasserre hierarchy has an integrality gap $\approx 2$ for Max 3-LIN [Sch08]. Via reductions from this result, Tulsiani showed gap instances for Max $k$-CSP (for $\Omega(n)$ rounds), and instances with $n^{1-o(1)}$ gap for $\approx 2^{\sqrt{\log n}}$ rounds for the Independent Set and Chromatic Numbers [Tul09].

---

[1]In fact our work [GS11] was motivated by the analysis in this paper.
[2]It is known that $\lambda_{r \cdot kn^\delta}(\mathcal{L}(\widehat{G})) \geqslant \delta \lambda_r(\mathcal{L}(G))$ [ABS10], but this large multiplicative $n^\delta$ slack makes this ineffective for $r = n^{o(1)}$.

In terms of algorithmic results, even few rounds of Lasserre is already as strong as the SDPs used to obtain the best known approximation algorithms for several problems — for example, 3 rounds of Lasserre is enough to capture the ARV SDP relaxation for Sparsest Cut [ARV09], and Chlamtac used the third level of the Lasserre hierarchy to get improvements for coloring 3-colorable graphs [Chl07]. In terms of positive results that use a larger (growing) number of Lasserre rounds, we are aware of only two results. Chlamtac and Singh used $O(1/\gamma^2)$ rounds of Lasserre hierarchy to find an independent set of size $\Omega(n^{\gamma^2/8})$ in 3-uniform hypergraphs with an independent set of size $\gamma n$ [CS08]. Karlin, Mathieu, and Nguyen show that $1/\varepsilon$ rounds of Lasserre SDP gives a $(1+\varepsilon)$ approximation to the Knapsack problem [KMN10].

However, there are *mixed* hierarchies, which are weaker than Lasserre and based on combining an LP characterized by local distributions (from the Sherali-Adams hierarchy) with a simple SDP, that have been used for several approximation algorithms. For instance, for the above-mentioned result on independent sets in 3-uniform hypergraphs, an $n^{\Omega(\gamma^2)}$ sized independent set can be found with $O(1/\gamma^2)$ levels from the mixed hierarchy. Raghavendra's result states that for every constraint satisfaction problem, assuming the Unique Games conjecture, the best approximation ratio is achieved by a small number of levels from the mixed hierarchy [Rag08]. For further information and references on the use of SDP and LP hierarchies in approximation algorithms, we point the reader to the excellent book chapter [CT11].

In an independent work, Barak, Raghavendra, and Steurer [BRS11] consider the above-mentioned mixed hierarchy, and extend the local propagation rounding of [AKK$^+$08] to these SDPs in a manner similar to our work. Their analysis methods are rather different from ours. Instead of column-based low-rank matrix approximation, they use the graph spectrum to infer global correlation amongst the SDP vectors from local correlation, and use it to iteratively to argue that a random seed set works well in the rounding. Their main result is an *additive approximation for Max 2-CSPs*. Translating to the terminology used in this paper, given a 2CSP instance over domain size $k$ with optimal value (fraction of satisfied constraints) equal to $v$, they give an algorithm to find an assignment with value $v - O\left(k\sqrt{1-\lambda_r}\right)$ based on $r' \gg kr$ rounds of the mixed hierarchy. (Here $\lambda_r$ is the $r$'th smallest eigenvalue of the normalized Laplacian of the *constraint* graph; note though that $\lambda_r$ needs to be fairly close to 1 for the bound to kick in.) For the special case of Unique Games, they get the better performance of $v - O\left(\sqrt[4]{1-\lambda_r}\right)$ which doesn't degrade with $k$, and also a factor $O(1/\lambda_r)$ approximation for minimizing the number of unsatisfied constraints in time *exponential* in $k$.

For 2CSPs, our results only apply to a restricted class (corresponding to PSD quadratic forms), but we get approximation-scheme style *multiplicative* guarantees for the harder *minimization objective*, and can handle *global linear constraints*. (Also, for Unique Games, our algorithm has running time *polynomial* in the number of labels $k$. In terms of $r$, the runtime in [BRS11] has a better $2^{O(r)}$ type dependence instead of our $n^{O(r)}$ bounds.) Our approach enables us to get approximation-scheme style guarantees for several notorious graph partitioning problems that have eluded even APX-hardness.

## 1.4 Organization

We begin with the definition of Lasserre SDPs we will use in Section 2. To illustrate the main ideas in our work, we present them in a self-contained way for a simplified setting in Section 3, by developing an algorithm for the Minimum Bisection problem.

The rest of our results are proved in remaining Sections: quadratic integer programming in Section 4; graph partitioning problems such as conductance, Uniform Sparsest Cut, and SSE (on general, non-regular, weighted graphs) in Section 5; Unique Games in Section 6; and finally, independent sets in graphs in Section 8.

The main technical theorem about rounding that is used to analyze our algorithm for quadratic integer programming is proved in Section 7. We discuss the accuracy needed (by our rounding) in the solution to the Lasserre SDP in Appendix A.

## 2 Lasserre hierarchy of semidefinite programs

We present the formal definitions of the Lasserre family of SDP relaxations [Las02], tailored to the setting of the problems we are interested in, where the goal is to assign to each vertex/variable from a set $V$ a label from $[k] = \{1, 2, \ldots, k\}$.

**Definition 1** (Lasserre vector set). *Given a set of variables $V$ and a set $[k] = \{1, 2, \ldots, k\}$ of labels, and an integer $r \geqslant 0$, a vector set $x$ is said to satisfy $r$-levels of Lasserre hierarchy constraints on $k$ labels, denoted*
$$x \in \mathbf{Lasserre}^{(r)}(V \times [k]) \,,$$

*if it satisfies the following conditions:*

1. *For each set $S \in \binom{V}{\leqslant r+1}$, there exists a function $x_S : [k]^S \to \mathbb{R}^\Upsilon$ that associates a vector of some finite dimension $\Upsilon$ with each possible labeling of $S$. We use $x_S(f)$ to denote the vector associated with the labeling $f \in [k]^S$. For singletons $u \in V$, we will use $x_u(i)$ and $x_u(i^u)$ for $i \in [k]$ interchangeably.*

   *For $f \in [k]^S$ and $v \in S$, we use $f(v)$ as the label $v$ receives from $f$. Also given sets $S$ with labeling $f \in [k]^S$ and $T$ with labeling $g \in [k]^T$ such that $f$ and $g$ agree on $S \cap T$, we use $f \circ g$ to denote the labeling of $S \cup T$ consistent with $f$ and $g$: If $u \in S$, $(f \circ g)(u) = f(u)$ and vice versa.*

2. $\|x_\emptyset\|^2 = 1$.
3. $\langle x_S(f), x_T(g) \rangle = 0$ *if there exists $u \in S \cap T$ such that $f(u) \neq g(u)$.*
4. $\langle x_S(f), x_T(g) \rangle = \langle x_A(f'), x_B(g') \rangle$ *if $S \cup T = A \cup B$ and $f \circ g = f' \circ g'$.*
5. *For any $u \in V$, $\sum_{j \in [k]} \|x_u(j)\|^2 = \|x_\emptyset\|^2$.*
6. *(implied by above constraints) For any $S \in \binom{V}{\leqslant r+1}$, $u \in S$ and $f \in [k]^{S \setminus \{u\}}$, $\sum_{g \in [k]^u} x_S(f \circ g) = x_{S \setminus \{u\}}(f)$.*

*We will use $\mathcal{X}(i)$ to denote a matrix of size $\Upsilon \times n$, $\mathcal{X}(i) \in \mathbb{R}^{\Upsilon \times V}$ whose columns are the vectors $\{x_u(i)\}_{u \in V}$.*

We now add linear constraints to the SDP formulation.

**Definition 2** (Linear constraints in Lasserre SDPs). *Given a matrix $B = [b_1 \ \ldots \ b_\ell] \in \mathbb{R}^{(V \times [k]) \times \ell}$ and a vector $c = (c_1, \ldots, c_\ell)^T \in \mathbb{R}^\ell$, $x \in \mathbf{Lasserre}^{(r)}(V \times [k])$, is said to satisfy linear constraints $\{(b_i, c_i)\}_{i=1}^\ell$ if the following holds for all $i \in [\ell]$:*

*For all subsets $S \in \binom{V}{\leqslant r}$ and $f \in [k]^V$,*

$$\sum_{u \in V, g \in [k]^u} \langle x_S(f), x_u(g) \rangle b_i(u, g) \leqslant c_i \langle x_S(f), x_\emptyset \rangle ,$$

*which is equivalent to*

$$\sum_{u \in V, g \in [k]^u} \|x_{S \cup \{u\}}(f \circ g)\|^2 b_i(u, g) \leqslant c_i \|x_S(f)\|^2.$$

*We denote the set of such $x$ as $x \in \mathbf{Lasserre}^{(r)}(V \times [k], B^{\leqslant c})$.*

**Remark 4** (Convenient matrix notation). One common expression we will use throughout this paper is the following. For matrices $X \in \mathbb{R}^{\Upsilon \times V}$ and $M \in \mathbb{R}^{V \times V}$:

$$\mathsf{Tr}(X^T X M) = \sum_{u,v \in V} M_{u,v} \langle X_u, X_v \rangle .$$

Note that if $M$ is positive semidefinite (denoted $M \succeq 0$), then $\mathsf{Tr}(X^T X M) \geqslant 0$.[3]

Also, if $L$ is Laplacian matrix of an undirected graph $G = (V, E)$, we have

$$\mathsf{Tr}(X^T X L) = \sum_{e = \{u,v\} \in E} \|X_u - X_v\|^2$$

where $X_u$ denotes the column of $X$ corresponding to $u \in V$. □

The analysis of our rounding algorithm will involve projections on certain subspaces, which we define next.

**Definition 3** (Projection operators). *Given $x \in \mathbf{Lasserre}^{(r)}(V \times [k])$, we define $\Pi : \binom{V}{\leqslant r+1} \to \mathbb{R}^{\Upsilon \times \Upsilon}$ as the projection matrix onto the span of $\{x_S(f)\}_{f \in [k]^S}$ for given $S$:*

$$\Pi_S \triangleq \sum_{f \in [k]^S} \overline{x_S(f)} \cdot \overline{x_S(f)}^T .$$

*(Here $\overline{x_S(f)}$ is the unit vector in the direction of $x_S(f)$ if $x_S(f)$ is nonzero, and $0$ otherwise.)*

*Similarly we define $P : \binom{V}{\leqslant r+1} \to \mathbb{R}^{\Upsilon \times \Upsilon}$ as the matrix corresponding to projection onto the span of $\{x_v(f)\}_{v \in S, f \in [k]}$: $P_S \triangleq \sum_{v \in S, f \in [k]} \overline{x_v(f)} \cdot \overline{x_v(f)}^T$.*

*We will denote by $\Pi_S^\perp = I - \Pi_S$ and $P_S^\perp = I - P_S$ the projection matrices onto the respective orthogonal complements, where $I$ denotes the identity matrix of appropriate dimension.*

**Remark 5** (Errors in Lasserre solution). In Appendix A, we present a simple algorithm which finds an SDP solution that satisfies all Lasserre consistency constraints *exactly*, with a small additive error in the linear constraints and the objective value. □

---
[3] The use of this inequality in various places is the reason why our analysis only works for minimizing PSD quadratic forms.

## 3 Case Study: Approximating Minimum Bisection

All our algorithmic results follow a unified method (except small set expansion on irregular graphs and unique games, both of which we treat separately). In this section, we will illustrate the main ideas involved in our work in a simplified setting, by working out progressively better approximation ratios for the following basic, well-studied problem: Given as input a graph $G = (V, E)$ and an integer size parameter $\mu$, find a subset $U \subset V$ with $|U| = \mu$ that minimizes the number of edges between $U$ and $V \setminus U$, denoted $\Gamma_G(U)$. The special case when $\mu = |V|/2$ and we want to partition the vertex set into two equal parts is the minimum bisection problem. We will loosely refer to the general $\mu$ case also as minimum bisection.[4]

For simplicity we will assume $G$ is unweighted and $d$-regular, however all our results given in Section 5 are for any weighted undirected graph $G$. We can formulate this problem as a binary integer programming problem as follows:

$$\min_{\widetilde{x}} \sum_{e=\{u,v\}\in E} (\widetilde{x}_u(1) - \widetilde{x}_v(1))^2, \tag{1}$$

$$\text{subject to} \sum_u \widetilde{x}_u(1) = \mu; \quad \forall u, \widetilde{x}_u(1) + \widetilde{x}_u(2) = 1; \quad \text{and} \quad \widetilde{x} \in \{0,1\}^{V \times [2]}.$$

If we let $L$ be the Laplacian matrix for $G$, we can rewrite the objective as $\eta \triangleq \widetilde{x}(1)^T L \widetilde{x}(1)$. We will denote by $\mathcal{L} = \frac{1}{d}L$ the normalized Laplacian of $G$.

Note that the above is a quadratic integer programming (QIP) problem with linear constraints. The somewhat peculiar formulation is in anticipation of the Lasserre semidefinite programming relaxation for this problem, which we describe below.

### 3.1 Lasserre relaxation for Minimum Bisection

Let $b$ be the vector on $V \times [2]$ with $b_v(1) = 1$ and $b_v(2) = 0$ for every $v \in V$. For an integer $r' \geq 0$, the $r'$-round Lasserre SDP for Minimum Bisection consists of finding $x \in \mathbf{Lasserre}^{(r')}(V \times [k], b^{=\mu})$ that minimizes the objective function

$$\sum_{e=\{u,v\}\in E(G)} \|x_u(1) - x_v(1)\|^2 . \tag{2}$$

It is easy to see that this is indeed a relaxation of our original QIP formulation (1).

### 3.2 Main theorem on rounding

Let $x$ be an (optimal) solution to the above $r'$-round Lasserre SDP. We will always use $\eta$ in this section to refer to the objective value of $x$, i.e., $\eta = \sum_{e=\{u,v\}\in E(G)} \|x_u(1) - x_v(1)\|^2$.

Our ultimate goal in this section is to give an algorithm to round the SDP solution $x$ to a good cut $U$ of size very close to $\mu$, and prove the below theorem.

---

[4]We will be interested in finding a set of size $\mu \pm o(\mu)$, so we avoid the terminology *Balanced Separator* which typically refers to the variant where $\Omega(n)$ slack is allowed in the set size.

**Theorem 4.** *For all $r \geqslant 1$ and $\varepsilon > 0$, there exists $r' = O\left(\frac{r}{\varepsilon^2}\right)$, such that given $x \in \mathbf{Lasserre}^{(r')}(V \times [k], b^{=\mu})$ with objective value (2) equal to $\eta$, one can find in randomized $n^{O(1)}$ time, a set $U \subseteq V$ satisfying the following two properties w.h.p:*

1. $\Gamma_G(U) \leqslant \frac{1+\varepsilon}{\min(1,\lambda_{r+1}(\mathcal{L}))}\eta.$
2. $\mu(1 - o(1)) = \mu - O\left(\sqrt{\mu \log(1/\varepsilon)}\right) \leqslant |U| \leqslant \mu + O\left(\sqrt{\mu \log(1/\varepsilon)}\right) = \mu(1 + o(1)).$

Since one can solve the Lasserre relaxation in $n^{O(r')}$ time using Theorem 39, we get the result claimed in the introduction: an $n^{O(r/\varepsilon^2)}$ time factor $(1+\varepsilon)/\min\{\lambda_r, 1\}$ approximation algorithm; the formal theorem, for general (non-regular, weighted) graphs appears as Corollary 15 in Section 5. Note that if $t = \arg\min_r\{r \mid \lambda_r(\mathcal{L}) \geqslant 1 - \varepsilon/2\}$, then this gives an $n^{O_\varepsilon(t)}$ time algorithm for approximating minimum bisection to within a $(1+\varepsilon)$ factor, provided we allow $O(\sqrt{n})$ imbalance.

### 3.3 The rounding algorithm

Recall that the solution $x \in \mathbf{Lasserre}^{(r')}(V \times [k], b^{=\mu})$ contains a vector $x_T(f)$ for each $T \in \binom{V}{\leqslant r'}$ and every possible labeling of $T$, $f \in [2]^T$ of $T$. Our approach to round $x$ to a solution $\widetilde{x}$ to the integer program (1) is similar to the label propagation approach used in [AKK+08].

Consider fixing a set of $r'$ nodes, $S \in \binom{V}{r'}$, and assigning a label $f(s)$ to every $s \in S$ by choosing $f \in [2]^S$ with probability $\|x_S(f)\|^2$. (The best choice of $S$ can be found by brute-forcing over all of $\binom{V}{r'}$, since solving the Lasserre SDP takes $n^{O(r')}$ time anyway. But there is also a faster method to find a good $S$, as mentioned in Theorem 7.) Conditional on choosing a specific labeling $f$ to $S$, we propagate the labeling to other nodes as follows: Independently for each $u \in V$, choose $i \in [2]$ and assign $\widetilde{x}_u(i) \leftarrow 1$ with probability

$$\Pr[\widetilde{x}_u(i) = 1] = \frac{\|x_{S \cup \{u\}}(f \circ i^u)\|^2}{\|x_S(f)\|^2} = \frac{\langle \overline{x_S(f)}, x_u(i) \rangle}{\|x_S(f)\|}.$$

Observe that if $u \in S$, label of $u$ will always be $f(u)$. Finally, output $U = \{u \mid \widetilde{x}_u(1) = 1\}$ as the cut. Below $\Pi_S$ denotes the projection matrix from Definition 3.

**Lemma 5.** *For the above rounding procedure, the size of the cut produced $\Gamma_G(U)$ satisfies*

$$\mathbb{E}[\Gamma_G(U)] = \eta + \sum_{(u,v) \in E} \langle \Pi_S^\perp x_u(1), \Pi_S^\perp x_v(1) \rangle. \tag{3}$$

*Proof.* Note that for $u \neq v$, and $i, j \in [2]$,

$$\Pr[\widetilde{x}_u(i) = 1 \wedge \widetilde{x}_v(j) = 1] = \sum_f \|x_S(f)\|^2 \frac{\langle \overline{x_S(f)}, x_u(i) \rangle}{\|x_S(f)\|} \frac{\langle \overline{x_S(f)}, x_v(j) \rangle}{\|x_S(f)\|}$$

$$= \sum_f \langle \overline{x_S(f)}, x_u(i) \rangle \langle \overline{x_S(f)}, x_v(j) \rangle.$$

Since $\{\overline{x_S(f)}\}_f$ is an orthonormal basis, the above expression can be written as the inner product of *projections* of $x_u(i)$ and $x_v(j)$ onto the span of $\{x_S(f)\}_{f \in [2]^S}$, which we denote by $\Pi_S$. Let us now calculate the expected number $\Gamma_G(U)$ of edges cut by this rounding. It is slightly more convenient to treat edges $e = \{u, v\}$ as two directed edges $(u, v)$ and $(v, u)$, and count directed edges $(u, v)$ with $u \in U$ and $v \in V \setminus U$ in the cut. Therefore,

$$\mathbb{E}\left[\Gamma_G(U)\right] = \sum_{(u,v) \in E} \langle \Pi_S x_u(1), \Pi_S x_v(2) \rangle = \sum_{(u,v) \in E} \langle \Pi_S x_u(1), \Pi_S(x_\emptyset - x_v(1)) \rangle$$
$$= \sum_{(u,v) \in E} \langle \Pi_S x_u(1), \Pi_S x_\emptyset \rangle - \langle \Pi_S x_u(1), \Pi_S x_v(1) \rangle \quad (4)$$

By using the fact that $\langle \Pi_S x_u(1), \Pi_S x_\emptyset \rangle = \langle x_u(1), \Pi_S x_\emptyset \rangle = \langle x_u(1), x_\emptyset \rangle = \|x_u(1)\|^2$, we can rewrite Equation (4) in the following way:

$$= \sum_{(u,v) \in E} \|x_u(1)\|^2 - \langle \Pi_S x_u(1), \Pi_S x_v(1) \rangle$$
$$= \sum_{(u,v) \in E} \|x_u(1)\|^2 - \langle x_u(1), x_v(1) \rangle + \langle \Pi_S^\perp x_u(1), \Pi_S^\perp x_v(1) \rangle$$
$$= \eta + \sum_{(u,v) \in E} \langle \Pi_S^\perp x_u(1), \Pi_S^\perp x_v(1) \rangle. \qquad \square$$

Note that the matrix $\Pi_S$ depends on vectors $x_S(f)$ which are hard to control because we do not have any constraint relating $x_S(f)$ to a known matrix. The main driving force behind all our results is the following fact, which follows since given any $u \in S$ and $i \in [2]$, $x_u(i) = \sum_{f:f(u)=i} x_S(f)$ by Lasserre constraints.

**Observation 6.** *For all $S \in \binom{V}{r'}$,*

$$\mathrm{span}\left(\{x_S(f)\}_{f \in [2]^S}\right) \supseteq \mathrm{span}\left(\{x_u(i)\}_{u \in S, i \in [2]}\right) .$$

*Equivalently for $P_S$ being the projection matrix onto span of $\{x_u(i)\}_{u \in S, i \in [2]}$, $P_S \preceq \Pi_S$.*

Thus we will try to upper bound the term in Equation (3) by replacing $\Pi_S^\perp$ with $P_S^\perp$, but we cannot directly perform this switch: $\langle P_S^\perp x_u(i), P_S^\perp x_v(j) \rangle$ might be negative while $\Pi_S^\perp x_u(i) = 0$.

### 3.4 Factor $1 + \frac{1}{\lambda_r}$ approximation of cut value

Our first bound is by directly upper bounding Equation (3) in terms of $\|\Pi_S^\perp x_u(i)\|^2 \leqslant \|P_S^\perp x_u(i)\|^2$. Using Cauchy-Schwarz and Arithmetic-Geometric Mean inequalities, (3) implies that the expected number of edges cut is upper bounded by

$$\eta + \frac{1}{2} \sum_{e=(u,v) \in E} \|\Pi_S^\perp x_u(1)\|^2 + \|\Pi_S^\perp x_v(1)\|^2 = \eta + d \sum_u \|\Pi_S^\perp x_u(1)\|^2 \leqslant \eta + d \sum_u \|P_S^\perp x_u(1)\|^2 . \quad (5)$$

Now define $X_u \triangleq x_u(1)$, and let $X \in \mathbb{R}^{\Upsilon \times V}$ be the matrix with columns $X_u$. By (2), we have the objective value $\eta = \mathsf{Tr}(X^T X L)$. Let $X_S^\Pi \triangleq \sum_{u \in S} \overline{X_u X_u}^T$ be the projection matrix onto the span of $\{X_u\}_{u \in S}$. Since this set is a subset of $\{x_u(i)\}_{u \in S, i \in [2]}$, we have $X_S^\Pi \preceq P_S$. Therefore, we can bound (5) further as

$$\mathbb{E}\left[\text{number of edges cut}\right] \leq \eta + d \sum_u \|X_S^\perp X_u\|^2 = \eta + d \cdot \text{Tr}(X^T X_S^\perp X) \,. \tag{6}$$

To get the best upper bound, we want to pick $S \in \binom{V}{r'}$ to minimize $\sum_{u \in V} \|X_S^\perp X_u\|^2$. It is a well known fact that among *all* projection matrices $M$ of rank $r'$ (not necessarily restricted to projection onto columns of $X$), the minimum value of $\sum_u \|M^\perp X_u\|^2 = \text{Tr}(X^T M^\perp X)$ is achieved by matrix $M$ projecting onto the space of the largest $r'$ singular vectors of $X$. Further, this minimum value equals $\sum_{i \geq r'+1} \sigma_i$ where $\sigma_i = \sigma_i(X)$ denotes the *squared $i^{th}$* largest singular value of $X$ (equivalently $\sigma_i(X)$ is the $i^{th}$ largest eigenvalue of $X^T X$). Hence $\text{Tr}(X^T X_S^\perp X) \geq \sum_{i \geq r'+1} \sigma_i$ for every choice of $S$. The following theorem from [GS11] shows the existence of $S$ which comes close to this lower bound:

**Theorem 7.** *[GS11] For every real matrix $X$ with column set $V$, and positive integers $r \leq r'$, we have*

$$\delta_{r'}(X) \triangleq \min_{S \in \binom{V}{r'}} \text{Tr}(X^T X_S^\perp X) \leq \frac{r'+1}{r'-r+1}\left(\sum_{i \geq r+1} \sigma_i\right) \,.$$

*In particular, for all $\varepsilon \in (0, 1)$, $\delta_{r/\varepsilon} \leq \frac{1}{1-\varepsilon}\left(\sum_{i \geq r+1} \sigma_i\right)$. Further one can find a set $S \in \binom{V}{r'}$ achieving the claimed bounds in deterministic $O(rn^4)$ time.*

**Remark 6.** Prior to our paper [GS11], it was shown in [BDMI11] that $\delta_{r(2+\varepsilon)/\varepsilon} \leq (1+\varepsilon)\left(\sum_{i \geq r+1} \sigma_i\right)$. The improvement in the bound on $r'$ from $2r/\varepsilon$ to $r/\varepsilon$ to achieve $(1 + \varepsilon)$ approximation is not of major significance to our application, but since the tight bound is now available, we decided to state and use it. □

**Remark 7** (Running time of our algorithms). If the Lasserre SDP can be solved faster than $n^{O(r')}$ time, perhaps in $\exp(O(r'))n^c$ time for some absolute constant $c$, then the fact that we can find $S$ deterministically in only $O(n^5)$ time would lead to a similar runtime for the overall algorithm. □

Picking the subset $\mathcal{S}^* \in \binom{V}{r'}$ that achieves the bound (6) guaranteed by Theorem 7, we have

$$\text{Tr}(X^T X_{\mathcal{S}^*}^\perp X) = \delta_{\frac{r}{\varepsilon}}(X) \leq (1-\varepsilon)^{-1} \sum_{i > r} \sigma_i \,.$$

In order to relate this quantity to the SDP objective value $\eta = \text{Tr}(X^T X L)$, we use the fact that $\text{Tr}(X^T X L)$ is minimized when eigenvectors of $X^T X$ and $L$ are matched in reverse order: $i^{th}$ largest eigenvector of $X^T X$ corresponds to $i^{th}$ smallest eigenvector of $L$. Letting $0 = \lambda_1(\mathcal{L}) \leq \lambda_2(\mathcal{L}) \leq \ldots \leq \lambda_n(\mathcal{L}) \leq 2$ be the eigenvalues of normalized graph Laplacian matrix, $\mathcal{L} = \frac{1}{d}L$, we have

$$\frac{\eta}{d} = \frac{1}{d}\text{Tr}(X^T X L) \geq \sum_i \sigma_i(X)\lambda_i(\mathcal{L}) \geq \sum_{i \geq r+1} \sigma_i(X)\lambda_{r+1}(\mathcal{L}) \geq (1-\varepsilon)\lambda_{r+1}(\mathcal{L})\delta_{\frac{r}{\varepsilon}}(X).$$

Plugging this into (6), we can conclude our first bound:

**Theorem 8.** *For all positive integers $r$ and $\varepsilon \in (0,1)$, given SDP solution $x \in$ **Lasserre**$^{(\lceil r/\varepsilon \rceil)}(V \times [k], b^{=\mu})$, the rounding algorithm given in 3.3 cuts at most*

$$\left(1 + \frac{1}{(1-\varepsilon)\lambda_{r+1}(\mathcal{L})}\right) \sum_{e=(u,v)\in E} \|x_u(1) - x_v(1)\|^2$$

*edges in expectation.*

*In particular, the algorithm cuts at most a factor $\left(1 + \frac{1}{(1-\varepsilon)\lambda_{r+1}(\mathcal{L})}\right)$ more edges than the optimal cut with $\mu$ nodes on one side.*[5]

Note that $\lambda_n(\mathcal{L}) \leq 2$, hence even if we use $n$-rounds of Lasserre relaxation, for which $x$ is an integral solution, we can only show an upper bound $\geq \frac{3}{2}$. Although this is too weak by itself for our purposes, this bound will be crucial to obtain our final bound.

## 3.5 Improved analysis and factor $\frac{1}{\lambda_r}$ approximation on cut value

First notice that Equation (3) can be written as

$$\mathbb{E}[\text{number of edges cut}] = \text{Tr}(X^T X L) + \text{Tr}(X^T \Pi_S^\perp X(I-L)) = \text{Tr}(X^T \Pi_S^\perp X) + \text{Tr}(X^T \Pi_S X L). \quad (7)$$

If value of this expression is larger than $\frac{\eta}{(1-\varepsilon)\lambda_{r+1}} + \eta\varepsilon$, then value of $\text{Tr}(X^T \Pi_S X L)$ has to be larger than $\varepsilon\eta$ due to the bound we proved on $\text{Tr}(X^T \Pi_S^\perp X)$. Consider choosing another subset $T$ that achieves the bound $\delta_r(\Pi_S^\perp X)$. The crucial observation is that distances between neighboring nodes on vectors $\Pi_S^\perp X$ has decreased by an additive factor of $\eta\varepsilon$,

$$\text{Tr}(X^T \Pi_S^\perp X L) = \text{Tr}(X^T X L) - \text{Tr}(X^T \Pi_S X L) < \eta(1-\varepsilon)$$

so that $\text{Tr}(X^T \Pi_{S\cup T}^\perp X) < (1-\varepsilon)\frac{\eta}{(1-\varepsilon)\lambda_{r+1}}$. Now, if we run the rounding algorithm with $S \cup T$ as the seed set, and (7) with $S \cup T$ in place of $S$ is larger than $\frac{\eta}{(1-\varepsilon)\lambda_{r+1}} + \eta\varepsilon$, then $\text{Tr}(X^T \Pi_{S\cup T} X L) > 2\varepsilon\eta$. Hence

$$\text{Tr}(X^T \Pi_{S\cup T}^\perp X L) \leq \text{Tr}(X^T X L) - \text{Tr}(X^T \Pi_{S\cup T} X L) < \eta(1-2\varepsilon).$$

Picking another set $T'$, we will have $\text{Tr}(X^T \Pi_{S\cup T\cup T'}^\perp X) < (1-2\varepsilon)\frac{\eta}{(1-\varepsilon)\lambda_{r+1}}$. Continuing this process, if the quantity (7) is not upper bounded by $\frac{\eta}{(1-\varepsilon)\lambda_{r+1}} + \eta\varepsilon$ after $\lceil \frac{1}{\varepsilon} \rceil$ many such iterations, then the total projection distance becomes

$$\text{Tr}(X^T \Pi_{S\cup T\cup\ldots}^\perp X) < (1 - \lceil 1/\varepsilon \rceil \varepsilon)\frac{\eta}{(1-\varepsilon)\lambda_{r+1}} \leq 0$$

which is a contradiction. For formal statement and proof in a more general setting, see Theorem 31 in Section 7.

**Theorem 9.** *For all integers $r \geq 1$ and $\varepsilon \in (0,1)$, letting $r' = O\left(\frac{r}{\varepsilon^2}\right)$, given SDP solution $x \in$ **Lasserre**$^{(\lceil r' \rceil)}(V \times [k], b^{=\mu})$, the expected number of edges cut by the above rounding algorithm is at most $(1+\varepsilon)/\min\{1, \lambda_{r+1}(\mathcal{L})\}$ times the size of the optimal cut with $\mu$ nodes on one side. (Here $\lambda_{r+1}(\mathcal{L})$ is the $(r+1)$'th smallest eigenvalue of the normalized Laplacian $\mathcal{L} = \frac{1}{d}L$ of the $G$.)*

---
[5]We will later argue that the cut will also meet the balance requirement up to $o(\mu)$ vertices.

## 3.6 Bounding Set Size

We now analyze the balance of the cut, and show that we can ensure that $|U| = \mu \pm o(\mu)$ in addition to $\Gamma_G(U)$ being close to the expected bound of Theorem 9 (and similarly for Theorem 8).

Let $\mathcal{S}^*$ fixed to be $\arg\min_{S \in \binom{V}{r'}} \mathsf{Tr}(X^T X_S^\perp X)$. We will show that conditioned on finding cuts with small $\Gamma_G(U)$, the probability that one of them has $|U| \approx \mu$ is bounded away from zero. We can use a simple Markov bound to show that there is a non-zero probability that both cut size and set size are within 3-factor of corresponding bounds. But by exploiting the independence in our rounding algorithm and Lasserre relaxations of linear constraints, we can do much better. Note that in the $r'$-round Lasserre relaxation, for each $f \in [2]^{\mathcal{S}^*}$, due to the set size constraint in original IP formulation, $x$ satisfies:

$$\sum_u \widetilde{x}_u(1) = \mu \implies \sum_u \langle x_{\mathcal{S}^*}(f), x_u(1) \rangle = \mu \|x_{\mathcal{S}^*}(f)\|^2 .$$

This implies that conditioned on the choice of $f$, the expectation of $\sum_u \widetilde{x}_u(1)$ is $\mu$ and events $\widetilde{x}_u(1) = 1$ for various $u$ are independent. Applying the Chernoff bound, we get

$$\Pr_{\widetilde{x}} \left[ \left| \sum_u \widetilde{x}_u(1) - \mu \right| \geq 2\sqrt{\mu \log \frac{1}{\zeta}} \right] \leq o(\zeta) \leq \frac{\zeta}{3}.$$

Consider choosing $f \in [2]^{\mathcal{S}^*}$ so that $\mathbb{E}\left[\text{number of edges cut} \mid f\right] \leq \mathbb{E}\left[\text{number of edges cut}\right] \triangleq b$. By Markov inequality, if we pick such an $f$, $\Pr\left[\text{number of edges cut} \geq (1+\zeta)b\right] \leq 1 - \frac{\zeta}{2}$, where the probability is over the random propagation once $\mathcal{S}^*$ and $f$ are fixed.

Hence with probability at least $\frac{\zeta}{6}$, the solution $\widetilde{x}$ will yield a cut $U$ with $\Gamma_G(U) \leq (1+\zeta)b$ and size $|U|$ in the range $\mu \pm 2\sqrt{\mu \log \frac{1}{\zeta}}$. Taking $\zeta = \varepsilon$ and repeating this procedure $O\left(\varepsilon^{-1} \log n\right)$ times, we get a high probability statement and finish our main Theorem 4 on minimum bisection.

## 4 Algorithm for Quadratic Integer Programming

First we will state couple of concentration inequalities in a form suitable for us.

**Definition 10.** *Given $a \in \mathbb{R}^n$, consider $n$ independent Bernoulli random variables, $X_i$ with $\Pr[X_i = 1] = p_i$ and $\Pr[X_i = 0] = 1 - p_i$ for some $p_i \in [0, 1]$. For any $0 < \varepsilon < 1$, we define $\Delta_\varepsilon(a, \mu)$ as the minimum value that satisfies*

$$\Pr\left[ \left| \sum_i a_i X_i - \mathbb{E}\left[\sum_i a_i X_i\right] \right| \geq \Delta_\varepsilon(a, p) \right] \leq \varepsilon.$$

*subject to*

$$\mathbb{E}\left[\sum_i a_i X_i\right] = \mu.$$

**Corollary 11.** *For given vector $a \in \mathbb{R}^n$ and real $\mu$, and the following bounds hold:*

1. For any $a$, $\Delta_\varepsilon(a, \mu) = O\left(\sqrt{\ln \frac{1}{\varepsilon}} \|a\|_2\right)$.

2. If $a$ is non-negative and $\|a\|_\infty \leqslant \frac{\mu}{\log(1/\varepsilon)}$ then $\Delta_\varepsilon(a, \mu) = O\left(\sqrt{\|a\|_\infty \mu \log \frac{1}{\varepsilon}}\right)$.

*Proof of item 1.* Follows from Hoeffding bound. □

*Proof of Item 2.* Follows from Chernoff bound. □

**Definition 12** (Generalized eigenvalues). *Given two PSD matrices $A, B \in \mathbb{R}^{m \times m} \succeq 0$, $\lambda$ is a generalized eigenvalue along with the corresponding generalized eigenvector $x$ provided that $Bx \neq 0$ and*
$$Ax = \lambda Bx.$$
*We will use $\lambda_i[A; B]$ to denote the $i^{th}$ smallest generalized eigenvalue.*

**Observation 13.** *Given a PSD matrix $A$, $\lambda_i[A; \operatorname{diag}(A)]$ is equal to the $i^{th}$ smallest eigenvalue of the matrix $\operatorname{diag}(A)^{-1/2} \cdot A \cdot \operatorname{diag}(A)^{-1/2}$ where we use the convention $0/0 = \infty$ and $0 \cdot \infty = 0$.*

The result we present here is a generalization of minimum bisection. Since its proof is extremely similar to the one presented for minimum bisection, we only provide a sketch, while deferring the formal claim about the rounding analysis, Theorem 31, to Section 7.

**Theorem 14.** *Consider a quadratic integer programming problem*
$$\begin{aligned}
\min \quad & \widetilde{x}^T A \widetilde{x} \\
\text{s.t} \quad & B\widetilde{x} \geqslant c \\
& \widetilde{x}_u(i) = 0 && \text{for all } (u, i) \in D, \\
& \sum_{i \in [k]} \widetilde{x}_u(i) = 1 && \text{for all } u \in V, \\
& \widetilde{x} \in \{0, 1\}^{V \times [k]},
\end{aligned}$$
*where $A$ is a PSD matrix $A \succeq 0$, $D \subseteq V \times [k]$, $B \in \mathbb{R}^{\ell \times (V \times [k])}, c \in \mathbb{R}^\ell$. Further suppose that for all $i \in [\ell]$ and $u \in V$,*
$$\left|\left\{j \in [k] \,\middle|\, b_i(u, j) \neq 0\right\}\right| \leqslant 1.$$
*Moreover let $W$ be a positive real $W > 0$ such that $\|B\|_{\max} \leqslant 1$, $\|B\|_{\min} \geqslant \frac{1}{W}$,[6] and $\operatorname{Opt}$ (optimum value) is bounded from below by $\frac{1}{W}$.*

*Given $0 < \varepsilon < 1$, positive integer $r$, there exists an algorithm, running in time $n^{O(r/\varepsilon^2)} O(\log W)$, to find a labeling $\widetilde{x}$ with objective value*
$$\widetilde{x}^T A \widetilde{x} \leqslant \frac{1 + \varepsilon}{\min\{1, \lambda_{r+1}\}} \operatorname{OPT},$$
*where $\lambda_{r+1} = \lambda_{r+1}[A; \operatorname{diag}(A)]$ is defined as in Definition 12.*

*Furthermore it satisfies the following properties:*

---
[6] Here $\|\cdot\|_{\max}$ and $\|\cdot\|_{\min}$ denotes the maximum and minimum non-zero entry in terms of absolute value respectively.

1. For all $u \in V$:
$$\sum_{i \in [k]} \widetilde{x}_u(i) = 1,$$

2. For all $(u, i) \in D$:
$$\widetilde{x}_u(i) = 0,$$

3. For each row of $B$, $b_i$,
$$\langle b_i, \widetilde{x} \rangle \geqslant c_i - \Delta_{\varepsilon/(2\ell)}(b_i, c_i).$$

*Proof.* We can compute a solution $x$ to $r'$-rounds of Lasserre relaxation of this problem, using Theorem 39 with $\varepsilon_0 = \left(\frac{1}{Wkn}\right)^{O(1)}$ and representing each constraint $\widetilde{x}_u(i) = 0$ as a monomial equality constraint. We know that $x$ has the following properties: $x \in \mathbf{Lasserre}^{(r')}(V \times [k], B^{\geqslant c - \frac{1}{W^{\Omega(1)}}})$ with objective value $x^T A x = \eta \leqslant \mathsf{Opt}(1 + o(\varepsilon))$, it obeys the hard constraints $x_u(i) = 0$ for $(u, i) \in D$. Given such $x$, let $\widetilde{x}$ be the rounded labeling promised by Theorem 31 (which obeys the first two properties by construction), with expected objective value
$$\mathbb{E}_{\widetilde{x}} \left[ \widetilde{x}^T A \widetilde{x} \right] \leqslant \eta \cdot \frac{1 + \varepsilon/2}{\min\{1, \lambda_{r+1}\}}.$$

Using Markov inequality,
$$\Pr_{\widetilde{x}} \left[ \widetilde{x}^T A \widetilde{x} \leqslant \eta \cdot \frac{1 + \varepsilon}{\min\{1, \lambda_{r+1}\}} \right] \geqslant 1 - \varepsilon.$$

For $i \in [\ell]$, probability that property 3 for $i^{th}$ linear constraint will not be satisfied is $\frac{\varepsilon}{2\ell}$ by definition of $\Delta_\varepsilon$. Taking union bound, we see that with probability $\Omega(\varepsilon)$, $\widetilde{x} \sim \mathcal{D}^*$ will satisfy all conditions and have $\widetilde{x}^T A \widetilde{x} \leqslant \frac{1+\varepsilon}{\min\{1,\lambda_{r+1}\}}$. Repeating $O(n)$ many times, we can turn this into a high probability statement.

For the running time bound, the rounding algorithm takes randomized $n^{O(1)}$ time and solving the Lasserre relaxation takes $(kn)^{O(r')}W = n^{O(r')}W$ time. □

## 5 Algorithms for Graph Partitioning

Our goal in this section is to give approximation schemes for Minimum Bisection, Small Set Expansion, and various cut problems minimizing ratio of edges crossing the cut by the size/volume of the partition, such as Uniform Sparsest Cut, edge expansion, and conductance. Our results apply to weighted, not necessarily regular, graphs. Except for Minimum Bisection, it will be important for our algorithm that we are able to handle disjoint (possibly empty) foreground and background sets $F$ and $B$ and find a non-expanding set $U$ satisfying $F \subseteq U \subseteq V \setminus B$. So we state more general results with these additional constraints.

Throughout the whole paper, when talking about graphs, we use the following convention:

| Symbol | Stands for |
|---|---|
| $G$ | a connected, undirected weighted graph with non-negative edge weights. |
| $V, E, W$ | nodes, edges and edge weights of $G = (V, E, W)$. |
| $n$ | number of nodes, $n = |V|$. |
| $w_e$ | weight of edge $e = \{u, v\} \in E$. |
| $d_u$ | (weighted) degree of node $u \in V$, $d_u = \sum_{e=\{u,v\}\in E} w_e$. |
| $d_{\max}$ | maximum degree, $d_{\max} = \max_{u \in V} d_u$. |
| $m$ | total degree, $m = \sum_u d_u = \mathrm{Vol}_G(V)$. |
| $\mathrm{Vol}_G(U)$ | total degrees of nodes in $U \subseteq V$, $\mathrm{Vol}_G(U) = \sum_{u \in U} d_u$. |
| $\Gamma_G(U)$ | sum of weights of edges in the cut $[U, V \setminus U]$, $\Gamma_G(U) = \sum_{e=\{u,v\}\in E: u\in U, v\notin U} w_e$. |
| $D$ | diagonal matrix of node degrees, $D_{u,u} = d_u$, $D_{u,v} = 0$ for $u \neq v$. |
| $A$ | adjacency matrix of $G$, $A_{u,v} = w_{(u,v)}$ if $(u,v) \in E$, 0 otherwise. |
| $L$ | Laplacian matrix of $G$, $L = D - A$ |
| $\mathcal{L}$ | normalized Laplacian matrix of $G$, $\mathcal{L} = D^{-1/2} L D^{-1/2}$. |
| $\mathcal{A}$ | normalized adjacency matrix of $G$, $\mathcal{A} = D^{-1/2} A D^{-1/2}$. |
| $\lambda_i(\cdot)$ | $i^{th}$ smallest eigenvalue of given matrix. |
| OPT | optimum value (in terms of minimization) for the problem in question. |

For sake of clarity, let us formally recap the objectives of each of the cut problems we will solve in this section. In all cases, the input consists of a graph $G = (V, W, E)$, and disjoint foreground and background sets $F, B \subset V$.

- MINIMUM BISECTION: Given an integer $\mu$ satisfying $|V \setminus (F \cup B)| \leqslant \mu \leqslant |V|/2$, the objective function is the minimum value of $\Gamma_G(U)$ over sets $U$ such that $F \subseteq U \subseteq V \setminus B$ and $|U \setminus F| = \mu$.

- SMALL SET EXPANSION: Given an integer $\mu$ satisfying $\mathrm{Vol}_G(V \setminus (F \cup B)) \leqslant \mu \leqslant m/2$, the objective function is the minimum value of $\Gamma_G(U)$ over sets $U$ such that $F \subseteq U \subseteq V \setminus B$ and $\mathrm{Vol}_G(U \setminus F) = \mu$.[7]

- UNIFORM SPARSEST CUT: The objective function is to minimize
$$\phi_G(U) \triangleq \frac{\Gamma_G(U)}{|U| \cdot |V \setminus U|}.$$
over non-empty sets $U \subsetneq V$ such that $F \subseteq U \subseteq V \setminus B$.

- EDGE EXPANSION: The objective function is to minimize
$$h_G(U) \triangleq \frac{\Gamma_G(U)}{\min(|U|, |V \setminus U|)}.$$
over non-empty sets $U \subsetneq V$ such that $F \subseteq U \subseteq V \setminus B$.

- NORMALIZED CUT: The objective function is to minimize
$$\mathrm{ncut}_G(U) \triangleq \frac{\Gamma_G(U)}{\mathrm{Vol}_G(U) \cdot \mathrm{Vol}_G(V \setminus U)}.$$
over non-empty sets $U \subsetneq V$ such that $F \subseteq U \subseteq V \setminus B$.

---

[7]Our methods apply without change if the volume of the set $U \setminus F$ is only constrained to be in the range $[\mu(1 - \zeta), \mu(1 + \zeta)]$ for some $\zeta > 0$. For concreteness, we just focus on the exact volume case.

- CONDUCTANCE: The objective function is to minimize

$$\text{conductance}_G(U) \triangleq \frac{\Gamma_G(U)}{\min(\text{Vol}_G(U), \text{Vol}_G(V \setminus U))}.$$

over non-empty sets $U \subsetneq V$ such that $F \subseteq U \subseteq V \setminus B$.

## 5.1 Minimum Bisection

We will consider the following quadratic integer program formulation of Minimum Bisection

$$\min_{\widetilde{x}} \sum_{e=\{u,v\} \in E} (\widetilde{x}_u(1) - \widetilde{x}_v(1))^2,$$

$$\text{subject to } \sum_{u \in V \setminus F} \widetilde{x}_u(1) = \mu,$$

$$\widetilde{x}_u(1) = 1 \quad \forall u \in F,$$
$$\widetilde{x}_u(2) = 1 \quad \forall u \in B,$$
$$\widetilde{x}_u(1) + \widetilde{x}_u(2) = 1 \quad \forall u \in V,$$
$$\widetilde{x} \in \{0,1\}^{V \times [2]},$$

and its SDP relaxation:

$$\min_x \text{Tr}(\mathcal{X}(1)^T \mathcal{X}(1) L),$$

$$\text{subject to } \sum_{u \in V \setminus F} \|x_{S \cup \{u\}}(f \circ 1^u)\|^2 = \mu \|x_S(f)\|^2 \quad \forall S \in \binom{V}{r'} \text{ and } f \in [2]^S,$$

$$x_u(1) = x_\emptyset \quad \forall u \in F,$$
$$x_u(2) = x_\emptyset \quad \forall u \in B,$$
$$x \in \textbf{Lasserre}^{(r')}(V \times [2], b^{=\mu})$$

where $b$ is a vector with $b_u(1) = 1$ and $b_u(2) = 0$.

The first result is a straightforward corollary of Theorem 14.

**Corollary 15** (Minimum Bisection). *Given $0 < \varepsilon < 1$, positive integer $r$, a target size $\mu \leqslant \frac{n}{2}$, disjoint foreground and background sets (possibly empty) $F$ and $B$, respectively, with*

$$\mu \leqslant |V \setminus (F \cup B)|,$$

*there exists an algorithm, running in time $n^{O(r/\varepsilon^2)}$, to find a set $U \subseteq V$ such that:*

$$F \subseteq U \subseteq V \setminus B,$$

$$\mu - O\left(\sqrt{\mu \log(1/\varepsilon)}\right) \leqslant |U \setminus F| \leqslant \mu + O\left(\sqrt{\mu \log(1/\varepsilon)}\right),$$

$$\Gamma_G(U) \leqslant \frac{1+\varepsilon}{\min\{\lambda_{r+1}(\mathcal{L}), 1\}} \text{OPT}.$$

*Proof.* Follows from Theorem 14. Note that the objective matrix takes the form $L' = \begin{bmatrix} L & 0 \\ 0 & 0 \end{bmatrix}$. Its $(r+1)^{th}$ generalized smallest eigenvalue $\lambda_{r+1}[L'; \mathrm{diag}(L')]$ is equal to $(r+1)^{th}$ smallest eigenvalue $\lambda_{r+1}(\mathcal{L})$ of $\mathcal{L}$. □

## 5.2 Small Set Expansion

Our next result is on the small set expansion problem. A naïve application of Theorem 14 will yield good bounds only when the graph does not have high degree nodes (compared to the average degree). However our guarantee is irrespective of the degree distribution on graph $G$ such that we are always able to find a set of volume $\mu(1 \pm \varepsilon)$. In order to achieve this, the fact that we can assign arbitrary unary constraints in the form of foreground and background sets is crucial.

We use the following standard integer programming formulation of SSE

$$\min_{\widetilde{x}} \sum_{e=\{u,v\}\in E} (\widetilde{x}_u(1) - \widetilde{x}_v(1))^2,$$

$$\text{subject to } \sum_{u \in V \setminus F} d_u \widetilde{x}_u(1) = \mu,$$

$$\widetilde{x}_u(1) = 1 \quad \forall u \in F,$$

$$\widetilde{x}_u(2) = 1 \quad \forall u \in B,$$

$$\widetilde{x}_u(1) + \widetilde{x}_u(2) = 1 \quad \forall u \in V,$$

$$\widetilde{x} \in \{0,1\}^{V \times [2]},$$

and its natural SDP relaxation under Lasserre constraints:

$$\min_x \mathrm{Tr}(\mathcal{X}(1)^T \mathcal{X}(1) L),$$

$$\text{subject to } x_u(1) = x_\emptyset \quad \forall u \in F,$$

$$x_u(2) = x_\emptyset \quad \forall u \in B,$$

$$x \in \mathbf{Lasserre}^{(r')}(V \times [2], b^{=\mu})$$

where $b$ is a vector with $b_u(1) = d_u$ and $b_u(2) = 0$.

**Theorem 16** (Small Set Expansion). *Given $0 < \varepsilon < 1$, positive integer $r$, a target volume $\mu$, disjoint foreground and background sets (possibly empty) $F$ and $B$, respectively, with*

$$\mu \leqslant \mathrm{Vol}_G(V \setminus (F \cup B)),$$

*there exists an algorithm, running in time $n^{O\left(\frac{r+\log(1/\varepsilon)}{\varepsilon^2}\right)}$ to find a set $U \subseteq V$ such that:*

$$F \subseteq U \subseteq V \setminus B,$$

$$\Gamma_G(U) \leqslant \frac{1+\varepsilon}{\min\{\lambda_{r+1}(\mathcal{L}), 1\}} \mathsf{OPT}$$

*and*

1. If $d'_{\max} \triangleq \max_{u \in V \setminus (F \cup B)} d_u \leqslant O\left(\frac{\mu}{\log \frac{1}{\varepsilon}}\right)$, then

$$\mu \left(1 - O\left(\sqrt{\frac{d'_{\max}}{\mu} \log \frac{1}{\varepsilon}}\right)\right) \leqslant \mathrm{Vol}_G\left(U \setminus F\right) \leqslant \mu \left(1 + O\left(\sqrt{\frac{d'_{\max}}{\mu} \log \frac{1}{\varepsilon}}\right)\right)$$

   (In fact, in this case the running time is bounded by $n^{O(r/\varepsilon^2)}$.)

2. Else
$$\mu(1 - \varepsilon) \leqslant \mathrm{Vol}_G\left(U \setminus F\right) \leqslant \mu(1 + \varepsilon) .$$

*Proof of 1.* We follow the exact same proof as Theorem 14 using the standard integer programming formulation of SSE, and defining the cut $U = \{u \mid \widetilde{x}_u(1) = 1\}$. Using bound 2 from Corollary 11, $\Delta_{\varepsilon/2}(d, \mu) \leqslant O(\sqrt{d'_{\max} \mu \log(1/\varepsilon)})$ provided that $d'_{\max} \leqslant O\left(\frac{\mu}{\log \frac{1}{\varepsilon}}\right)$. Applying Markov bound on $\mathbb{E}_U[\Gamma_G(U)]$, we see that with probability $\Omega(\varepsilon)$,

$$\Gamma_G(U) \leqslant \frac{1 + \varepsilon}{\lambda_{r+1}}$$

and

$$\mu - O\left(\sqrt{\mu d'_{\max} \log(1/\varepsilon)}\right) \leqslant \mathrm{Vol}_G\left(U \setminus F\right) \leqslant \mu + O\left(\sqrt{\mu d'_{\max} \log(1/\varepsilon)}\right) . \qquad \square$$

*Proof of 2.* At a high level, our algorithm proceeds in the following way: We enumerate all subsets $U_0$ of volume at most $\mu$ from the set of high degree nodes $\mathcal{H}$, which is defined by $\mathcal{H} \triangleq \left\{u \mid d_u \geqslant \frac{\varepsilon^2}{\log(1/\varepsilon)} \mu\right\}$. For each such subset $U_0$, we solve the corresponding Lasserre SDP relaxation of Small Set Expansion problem on this graph with foreground set $F' \triangleq F \cup U_0$, background set $B' \triangleq B \cup (\mathcal{H} \setminus U_0)$ and target volume $\mu' \triangleq \mu - \mathrm{Vol}_G(U_0)$. Note that the maximum degree of any unconstrained node for this problem $d'''_{\max} = \max_{u \in V \setminus (F' \cup B')} d_u$ is at most $d'''_{\max} \leqslant \frac{\varepsilon^2}{\log(1/\varepsilon)} \mu$.

There are two possible cases:

1. If $d'''_{\max} < \frac{1}{\log(1/\varepsilon)} \mu'$ then we can apply the analysis given in proof of 1 to find a set $U$ such that
$$|\mathrm{Vol}_G(U_0 \cup U) - \mu| \leqslant O(\sqrt{\mu' d'''_{\max} \log(1/\varepsilon)}) \leqslant \sqrt{\mu' \varepsilon^2 \mu} \leqslant \sqrt{\mu \varepsilon^2 \mu} = \mu \varepsilon.$$

2. If $d'''_{\max} \geqslant \frac{1}{\log(1/\varepsilon)} \mu'$, then we have $\frac{1}{\log(1/\varepsilon)} \mu' \leqslant \frac{\varepsilon^2}{\log(1/\varepsilon)} \mu$, which implies $\mu' \leqslant \varepsilon^2 \mu$.

   In this case, instead of Chernoff bound, we use a simple Markov bound to conclude that

$$\mathrm{Pr}_U\left[\mathrm{Vol}_G(U) \geqslant \frac{\mu'}{\varepsilon/2}\right] \leqslant \varepsilon/2.$$

   Combining this with the bound on $\Gamma_G(U)$, with probability $\Omega(\varepsilon)$, we will find $U$ with

$$|\mathrm{Vol}_G(U_0 \cup U) - \mu| \leqslant \mu'/(\varepsilon/2) \leqslant 2\varepsilon\mu.$$

After enumerating all such sets, we return the one with smallest cut. Correctness of this algorithm is obvious.

For running time, note that number of nodes we can choose from $\mathcal{H}$ is at most $\frac{\log(1/\varepsilon)}{\varepsilon^2}$. Hence we solve Lasserre SDP at most
$$\binom{|\mathcal{H}|}{\leqslant \frac{\log(1/\varepsilon)}{\varepsilon^2}} \leqslant n^{O\left(\frac{\log(1/\varepsilon)}{\varepsilon^2}\right)}$$
times. Consequently the total running time is
$$n^{O\left(\frac{\log(1/\varepsilon)}{\varepsilon^2}\right)} \cdot n^{O\left(\frac{r}{\varepsilon^2}\right)} = n^{O\left(\frac{r+\log(1/\varepsilon)}{\varepsilon^2}\right)}.$$
□

## 5.3 Other Graph Partitioning Problems

Our final problems are graph partitioning problems with a ratio in the objective, uniform sparsest cut, edge expansion, normalized cut and conductance. All these results are direct extensions of our Minimum Bisection with arbitrary target sizes or Small Set Expansion results.

Note that the natural integer formulation for these problems involve a ratio in the objective and even after relaxing integrality constraints, the resulting formulation is not SDP anymore. Moreover due to the presence of Lasserre constraints, one can not simply equate the denominator to a constant, say 1, and solve the resulting SDP. We instead guess the value of denominator and solve the corresponding Min-Bisection or SSE problem, repeating this for all $\mathrm{poly}(n)$ possible values.

**Corollary 17** (Graph Partitioning). *Given $0 < \varepsilon < 1$, positive integer $r$, disjoint foreground and background sets (possibly empty) $F$ and $B$ respectively, there exists an algorithm, running in time $n^{O((r+\log(1/\varepsilon))/\varepsilon^2)}$, to find a non-empty set $U \subsetneq V$ such that:*
$$F \subseteq U \subseteq V \setminus B,$$
$$\{\phi_G(U), h_G(U), \mathrm{ncut}_G(U), \mathrm{conductance}_G(U)\} \leqslant \frac{1+\varepsilon}{\min\{\lambda_{r+1}(\mathcal{L}), 1\}} \mathtt{OPT}.$$

We sketch the proof for only conductance. Proofs for other problems follow the same pattern.

*Proof for Conductance.* Let $U^*$ be the optimal solution with $\mu^* = \mathrm{Vol}_G(U^*)$ and $\eta^* = \Gamma_G(U^*)$.

We guess the volume of optimal partition $\mu'$ and invoke Theorem 16. If we keep repeating this for all values of $\mu'$ in $\frac{\varepsilon m}{n}\{1, 2, \ldots, \lfloor\frac{n}{2\varepsilon}\rfloor\}$, we will find $U$ such that
$$|\mathrm{Vol}_G(U) - \mu^*| \leqslant \varepsilon \mu^* + o(1) \quad \text{and} \quad \Gamma_G(U) \leqslant \frac{1+\varepsilon}{\min\{1, \lambda_{r+1}\}}\eta^*$$
so that $\Phi_G(U) \leqslant \frac{1+O(\varepsilon)}{\min\{1,\lambda_{r+1}\}}\Phi_G(U^*)$. The running time will be bounded by
$$O\left(\frac{n}{\varepsilon}\right) \cdot n^{O\left(\frac{r+\log(1/\varepsilon)}{\varepsilon^2}\right)}.$$
□

## 5.4 Generalizations to $k$-way Partitioning Problems

Note that all these results can be generalized to their respective $k$-way partitioning versions (wherever it makes sense). The only difference is that, in each case, the objective matrix will be a block diagonal matrix consisting of $k$ copies of graph Laplacian matrix. It is easy to see that such a matrix has exactly $k$ copies of the original eigenvalues, so instead of $r$ rounds of Lasserre hierarchy, we will use $k \cdot r$ rounds instead.

**Corollary 18** (Minimum $k$-way Section). *Given $0 < \varepsilon < 1$, positive integer $r$ and a list of target set sizes $\{\mu_i\}_{i \in [k]}$ with $\sum_i \mu_i = n$, there exists an algorithm which runs in time $n^{O\left(\frac{kr}{\varepsilon^2}\right)}$ to find $k$ disjoint sets, $\{U_i\}_{i=1}^{k}$ such that:*

$$V = \bigsqcup_i U_i,$$

$$\forall i: \mu_i - O\left(\sqrt{\mu_i \log(k/\varepsilon)}\right) \leqslant |U_i| \leqslant \mu_i + O\left(\sqrt{\mu_i \log(k/\varepsilon)}\right),$$

$$\sum_i \Gamma_G(U_i) \leqslant \frac{1+\varepsilon}{\min\{\lambda_{r+1}(\mathcal{L}), 1\}} \mathsf{OPT}$$

*provided that such sets exist.*

*Here* $\mathsf{Opt}$ *is taken as minimum of $\sum_i \Gamma_G(U_i)$ over all $k$-way partitions of $V$ with $|U_i| = \mu_i$.*

*Proof.* Proof follows by applying Theorem 14 to:

$$\min_{\widetilde{x}} \sum_i \sum_{e=\{u,v\} \in E} w_e (\widetilde{x}_u(i) - \widetilde{x}_v(i))^2,$$

$$\text{subject to } \sum_{u \in V} \widetilde{x}_u(i) = \mu_i \quad \text{for all } i \in [k],$$

$$\sum_{i \in [k]} \widetilde{x}_u(i) = 1 \quad \forall u \in V,$$

$$\widetilde{x} \in \{0, 1\}^{V \times [k]}.$$

Let $\widehat{L}$ be the matrix in the objective. As remarked at the beginning of this section, the normalized matrix $\widehat{\mathcal{L}}$ has $k$ copies of each eigenvalue of $\mathcal{L}$, so $\widetilde{x}$ will satisfy

$$\widetilde{x}^T \widehat{L} \widetilde{x} \leqslant \frac{1 + O(\varepsilon)}{\min\{\lambda_{r/k}(\mathcal{L}), 1\}} \mathsf{Opt}$$

and set size constraints. Choosing $U_i \leftarrow \widetilde{x}^{-1}(i)$ completes the proof. □

## 6 Algorithms for Unique Games Type Problems

In this section, we obtain our algorithmic result for Unique Games type problems. Let us quickly recall the definition of the Unique Games problem. An instance of Unique Games consists of a graph $G = (V, E, W)$ with *non-negative edge weights* $w_e$ for each edge $e \in E$, a label set $[k]$, and bijection constraints $\pi_e : [k] \to [k]$ for each edge $e = \{u, v\}$. The goal is to find a labeling $f : V \to [k]$ that minimizes the number of unsatisfied constraints, where $e = \{u, v\}$ is unsatisfied if $\pi_e(f(u)) \neq f(v)$ (we assume the label of the lexicographically smaller vertex $u$ is projected by $\pi_e$).

**Remark 8.** Unique Games can also be captured in the quadratic integer programming framework of Section 4, where the matrix $A$ defining the PSD quadratic form corresponds to the Laplacian of the "lifted graph" $\widehat{G}$ with vertex set $V \times [k]$ obtained by replacing each edge in $G$ by a matching corresponding to its permutation constraint. However, except for the problem of maximum cut, we are unable to apply the results from that section directly because there is no known way to relate the $r^{th}$ eigenvalue of the constraint graph to say the $\text{poly}(r)^{th}$ eigenvalue of the lifted graph. Hence we use the "projection distance" type bound based on column selection (similar to Section 3.4), after constructing an appropriate embedding to relate the problem to the original graph. □

**Remark 9.** Although we do not explicitly mention in the theorem statements, we can provide similar guarantees in the presence of constraints similar to graph partitioning problems such as

- constraining labels available to each node,
- constraining fraction of labels used among different subsets of nodes.

For example, the guarantee for maximum cut algorithm immediately carries over to maximum bisection with guarantees on partition sizes similar to minimum bisection. □

## 6.1 Maximum cut

We first start with the simplest problem fitting in the framework for unique games — finding a maximum cut in a graph. We use the following standard integer programming formulation. Note that this formulation is for the complementary objective of finding minimum uncut:

$$\min_{\widetilde{x}} \sum_{e=\{u,v\} \in E} w_e \cdot \frac{1}{2} \left[(\widetilde{x}_u(1) - \widetilde{x}_v(2))^2 + (\widetilde{x}_u(2) - \widetilde{x}_v(1))^2\right],$$

$$\text{subject to } \widetilde{x}_u(1) + \widetilde{x}_u(2) = 1 \quad \forall u \in V,$$

$$\widetilde{x} \in \{0, 1\}^{V \times [2]},$$

and its natural SDP relaxation under Lasserre constraints:

$$\min_x \frac{1}{2} \text{Tr}\left[\begin{pmatrix} \mathcal{X}(1)^T \mathcal{X}(1) & \mathcal{X}(1)^T \mathcal{X}(2) \\ \mathcal{X}(2)^T \mathcal{X}(1) & \mathcal{X}(2)^T \mathcal{X}(2) \end{pmatrix} \begin{pmatrix} D & -A \\ -A^T & D \end{pmatrix}\right],$$

$$\text{subject to } x \in \textbf{Lasserre}^{(r')}(V \times [2]).$$

**Theorem 19** (Maximum Cut / Minimum Uncut). *Given a weighted undirected graph $G = (V, E, W)$, for all $\varepsilon \in (0, 1)$ and a positive integer $r$, there exists an algorithm to find a set $U \subseteq V$ such that the total weight of uncut edges by partitioning $(U, V \setminus U)$ is bounded by*

$$\min\left\{1 + \frac{2 + \varepsilon}{\lambda_{r+1}(\mathcal{L})}, \frac{1 + \varepsilon}{\min\{2 - \lambda_{n-r-1}(\mathcal{L}), 1\}}\right\} \cdot \texttt{OPT}$$

in time $n^{O(\frac{1}{\varepsilon})}2^{O(\frac{r}{\varepsilon^2})}$, where OPT *is the total weight of uncut edges in the optimal labeling.*

*Proof.* The first bound will follow from the more general result for Unique Games (Theorem 20 below), so we focus on the second bound claiming an approximation ratio of $(1+\varepsilon)/\min\{2-\lambda_{n-r-1},1\}$.

The Laplacian matrix, $\widehat{L}$ corresponding to the lifted graph, $\widehat{G}$, for maximum cut can be expressed as:

$$\widehat{L} = \begin{pmatrix} D & -A \\ -A^T & D \end{pmatrix} = \begin{pmatrix} D & -A \\ -A & D \end{pmatrix}$$

whose normalized Laplacian matrix is given by

$$\widehat{\mathcal{L}} = \begin{pmatrix} I & -\mathcal{A} \\ -\mathcal{A} & I \end{pmatrix}.$$

By direct substitution, it is easy to see that, for every eigenvector $q_i$ of constraint graph's normalized Laplacian matrix, $\mathcal{L}$, there are two corresponding eigenvectors for $\widehat{\mathcal{L}}$, $\begin{pmatrix} \frac{1}{\sqrt{2}} q_i \\ \frac{1}{\sqrt{2}} q_i \end{pmatrix}$ and $\begin{pmatrix} \frac{1}{\sqrt{2}} q_i \\ -\frac{1}{\sqrt{2}} q_i \end{pmatrix}$ with corresponding eigenvalues given by $\lambda_i$ and $2 - \lambda_i$ respectively. As a convention, we will refer to the first type of eigenvectors as even eigenvectors and the latter type as odd eigenvectors.

For any node $u \in V$, we can express $x_u(i)$ for $i \in [2]$ as

$$x_u(i) = \|x_u(i)\|^2 x_\emptyset + (-1)^i \|x_u(1)\| \|x_u(2)\| y_u \,,$$

where $y_u$ is a unit vector orthogonal to $x_\emptyset$, $\langle x_\emptyset, y_u \rangle = 0$. For any set $S$, $\Pi_S^\perp x_u(1) = \Pi_S^\perp(x_\emptyset^\perp x_u(1)) = \Pi_S^\perp y_u = -\Pi_S^\perp x_u(2)$. Consequently $\mathcal{X}^T \Pi_S^\perp \mathcal{X}$ has *zero* correlation with even eigenvectors of $\widehat{L}$. Therefore we have the following identity:

$$\mathsf{Tr}(\mathcal{X}^T \Pi_S^\perp x_\emptyset^\perp \Pi_S^\perp \mathcal{X} \widehat{L}) = \mathsf{Tr}(\mathcal{X}(1)^T \Pi_S^\perp x_\emptyset^\perp \Pi_S^\perp \mathcal{X}(1)(D+A)).$$

In particular, we can slightly modify Theorem 31 to take into account only the eigenvectors of $\widehat{\mathcal{L}}$ with which $x_\emptyset^\perp \mathcal{X}$ has non-zero correlation. Using our standard rounding procedure, we can then find a set $U$ for which the fraction of "uncut" edges is bounded by $(1+\varepsilon)\frac{\mathtt{OPT}}{\min(\lambda_{r+1}(I+\mathcal{A}),1)}$. The proof is now complete by noting that $\lambda_{r+1}(I + \mathcal{A}) = 2 - \lambda_{n-r-1}(\mathcal{L})$. □

## 6.2 Unique Games

In this section, we prove our main result for approximating Unique Games. We consider the following IP formulation:

$$\min_{\widetilde{x}} \sum_{e=\{u,v\} \in E} w_e \cdot \frac{1}{2} \sum_{i \in [k]} (\widetilde{x}_u(i) - \widetilde{x}_v(\pi_e(i)))^2,$$

$$\text{subject to } \sum_{i \in [k]} \widetilde{x}_u(i) = 1 \quad \forall u \in V,$$

$$\widetilde{x} \in \{0,1\}^{V \times [k]} \,,$$

and its natural SDP relaxation under Lasserre constraints:

$$\min_x \frac{1}{2} \text{Tr}\left(\mathcal{X}^T \mathcal{X} \hat{L}\right),$$
$$\text{subject to } x \in \textbf{Lasserre}^{(r')}(V \times [k]).$$

**Theorem 20** (Unique Games). *Let $G = (V, E, W)$ be an instance of Unique Games with label set $[k]$ and permutation constraints $\pi_e$ for each $e \in E$.*

*Then for all $\varepsilon \in (0, 1)$ and positive integer $r$, there exists an algorithm to find a labeling of nodes $V$, $f : V \to [k]$ with total weight of unsatisfied constraints bounded by*

$$\sum_{e=\{u,v\}\in E} w_e \mathbf{1}_{[f(v) \neq \pi_e(f(u))]} \leqslant \left(1 + \frac{2+\varepsilon}{\lambda_{r+1}(\mathcal{L})}\right) \texttt{OPT}$$

*in time $n^{O(1)} k^{O(\frac{r}{\varepsilon})}$, where* $\texttt{OPT}$ *is the total weight of unsatisfied constraints in the optimal labeling.*

*Proof.* Let $x$ be vectors satisfying $r' = O(r/\varepsilon)$-levels of Lasserre hierarchy constraints with

$$\eta \triangleq \frac{1}{4} \sum_{e=(u,v)\in E} w_e \sum_f \|x_u(f) - x_v(\pi_e(f))\|^2.$$

where for notational convenience we treat each undirected edge $\{u, v\}$ as two directed edges of half the weight. Let $S = \mathcal{S}^* \in \binom{V}{r'}$ to be chosen later. By choosing the labeling from $f \sim \mathcal{D}^*$, we know by using Claim 28 that the expected weight of unsatisfied constraints is bounded by:

$$\frac{1}{2} \sum_{e=(u,v)\in E} w_e \Pr_{f \sim \mathcal{D}^*}[f(u) \neq \pi_e(f(v))] = \frac{1}{2} \sum_{e=(u,v)\in E} w_e \sum_f \langle \Pi_S x_u(f), \Pi_S(x_\emptyset - x_v(\pi_e(f)))\rangle$$

$$= \frac{1}{2} \sum_{e=(u,v)\in E} w_e \sum_f \|x_u(f)\|^2 - \langle x_u(f), x_v(\pi_e(f))\rangle + \langle \Pi_S^\perp x_u(f), \Pi_S^\perp x_v(\pi_e(f))\rangle$$

$$= \eta + \frac{1}{2} \sum_{e=(u,v)\in E} w_e \sum_f \langle \Pi_S^\perp x_u(f), \Pi_S^\perp x_v(\pi_e(f))\rangle$$

$$\leqslant \eta + \frac{1}{2} \sum_{e=(u,v)\in E} w_e \sum_f \frac{\|\Pi_S^\perp x_u(f)\|^2 + \|\Pi_S^\perp x_v(f)\|^2}{2}$$

$$= \eta + \frac{1}{2} \sum_u d_u \sum_f \|\Pi_S^\perp x_u(f)\|^2$$

If we let $P_S$ be the projection matrix onto span of $\{x_v(f)\}_{v \in S, f \in [k]}$, the above is upper bounded by

$$\leqslant \eta + \frac{1}{2} \sum_u d_u \sum_f \|P_S^\perp x_u(f)\|^2.$$

Therefore the total weight of unsatisfied constraints is bounded by:

$$\eta^* \leqslant \eta \left(1 + \frac{\frac{1}{2}\sum_u d_u \sum_f \|P_S^\perp x_u(f)\|^2}{\frac{1}{4}\sum_{e=(u,v)\in E} \sum_f w_e \|x_u(f) - x_v(\pi_e(f))\|^2}.\right)$$

Consider the embedding $\{x_u(f)\}_{f\in[k]} \mapsto X_u$ given in Theorem 21 below. For this embedding, we know that, for any set $S$, the above quantity is bounded by

$$\leqslant \eta\left(1 + \frac{\frac{1}{2}\sum_u d_u \|X_S^\perp X_u\|^2}{\frac{1}{8}\sum_{e=(u,v)\in E} w_e \|X_u - X_v\|^2}\right) = \eta\left(1 + 4\frac{\mathsf{Tr}(X^T X_S^\perp XD)}{2\mathsf{Tr}(X^T XL)}\right)$$

If we further scale $X$ by $D^{1/2}$ so that $X' = D^{1/2}X$,

$$= \eta\left(1 + 2\frac{\mathsf{Tr}(X'^T X'^\perp_S X')}{\mathsf{Tr}(X'^T X'\mathcal{L})}\right)$$

where $\mathcal{L}$ is the normalized Laplacian matrix. Picking $\mathcal{S}^* = \arg\min_{C \in \binom{V}{r'}} \mathsf{Tr}(X'^T X'^\perp_C X')$, and applying Lemma 30 we obtain the desired result. $\square$

**Theorem 21** (A useful embedding). *Given vectors $x \in \mathbb{R}^{m \times (V \times [k])}$ with the property that, for any $u \in V$, whenever $f, g \in [k]^u$ are two different labellings of $u$, $f \neq g$,*

$$\langle x_u(f), x_u(g)\rangle = 0.$$

*Then there exists an embedding $\{x_u(f)\}_{f\in[k]^u} \mapsto X_u$ with the following properties:*

1. *For any $u \in V$, $\|X_u\|^2 = \sum_f \|x_u(f)\|^2$.*

2. *For any $u, v \in V$ and any permutation $\pi \in \mathrm{Sym}([k])$:*

$$\sum_{i\in[k]} \|x_u(i^u) - x_v(\pi(i)^v)\|^2 \geqslant \frac{1}{2}\|X_u - X_v\|^2.$$

3. *For any set $S \subseteq V$ and any node $u \in V$, if we let $P_S$ be the projection matrix onto the span of $\{x_s(f)\}_{s\in S, f\in[k]}$:*

$$\|X_S^\perp X_u\|^2 \geqslant \sum_{f\in[k]^u} \|P_S^\perp x_u(f)\|^2.$$

In the rest of this section, we will prove Theorem 21.

Our embedding is as follows. Assume that the vectors $x_u(f)$ belong to $\mathbb{R}^m$. Let $e_1, e_2, \ldots, e_m \in \mathbb{R}^m$ be the standard basis vectors. Define $X_u \in \mathbb{R}^m \otimes \mathbb{R}^m$ as

$$X_u = \sum_{i=1}^m \sum_{f\in[k]^u} \overline{\langle x_u(f), e_i\rangle} x_u(f) \otimes e_i .$$

**Observation 22.** *For vectors $x, y \in \mathbb{R}^m$, $\sum_{i=1}^m \langle x, e_i\rangle\langle y, e_i\rangle = \langle x, y\rangle$.*

The first property of the vectors $X_u$ follows from this observation easily:

$$\|X_u\|^2 = \sum_i \sum_{f,g} \langle \overline{x_u(f)}, e_i \rangle \langle \overline{x_u(g)}, e_i \rangle \langle x_u(f), x_u(g) \rangle$$

$$= \sum_{f,g} \langle x_u(f), x_u(g) \rangle \sum_i \langle \overline{x_u(f)}, e_i \rangle \langle \overline{x_u(g)}, e_i \rangle$$

$$= \sum_{f,g} \langle x_u(f), x_u(g) \rangle \langle \overline{x_u(f)}, \overline{x_u(g)} \rangle$$

$$= \sum_f \|x_u(f)\|^2.$$

We prove the second property in Claim 23 and third one in Claim 24.

**Claim 23.** *For any permutation $\pi \in \mathrm{Sym}([k])$:*

$$\frac{1}{2}\|X_u - X_v\|^2 \leqslant \sum_{i \in [k]} \|x_u(i^u) - x_v(\pi(i)^v)\|^2$$

*Proof.* Without loss of generality, we assume $\pi$ is the identity permutation. We have

$$\frac{1}{2}\|X_u - X_v\|^2 = \frac{\|X_u\|^2 + \|X_v\|^2}{2} - \langle X_u, X_v \rangle$$

$$= \frac{\|X_u\|^2 + \|X_v\|^2}{2} - \sum_{f,g} \langle x_u(f), x_v(g) \rangle \sum_i \langle \overline{x_u(f)}, e_i \rangle \langle \overline{x_v(g)}, e_i \rangle$$

$$= \sum_f \frac{\|x_u(f)\|^2 + \|x_v(f)\|^2}{2} - \sum_{f,g} \langle \overline{x_u(f)}, \overline{x_v(g)} \rangle^2 \|x_u(f)\| \|x_v(g)\|$$

The sum over all pairs is lower bounded by summing only the corresponding pairs:

$$\leqslant \frac{1}{2} \sum_f \left( \|x_u(f)\|^2 + \|x_v(f)\|^2 - 2\langle x_u(f), x_v(f) \rangle \langle \overline{x_u(f)}, \overline{x_v(f)} \rangle \right)$$

$$= \frac{1}{2} \sum_f \|x_u(f) - x_v(f)\|^2 + \sum_f \langle x_u(f), x_v(f) \rangle \underbrace{\left( 1 - \langle \overline{x_u(f)}, \overline{x_v(f)} \rangle \right)}_{\geqslant 0} \quad (8)$$

Since the coefficient of $\langle x_u(f), x_v(f) \rangle$ is positive, we can use Cauchy-Schwarz inequality to replace $\langle x_u(f), x_v(f) \rangle$ with $\|x_u(f)\| \cdot \|x_v(f)\|$ in Equation (8) to obtain:

$$\leqslant \frac{1}{2} \sum_f \|x_u(f) - x_v(f)\|^2 + \sum_f \left( \|x_u(f)\| \cdot \|x_v(f)\| - \langle x_u(f), x_v(f) \rangle \right) \quad (9)$$

Using inequality $\|x_u(f)\| \cdot \|x_v(f)\| \leqslant \frac{1}{2} \left( \|x_u(f)\|^2 + \|x_v(f)\|^2 \right)$ on Equation (9):

$$\leqslant \frac{1}{2} \sum_f \left( \|x_u(f) - x_v(f)\|^2 + \|x_u(f)\|^2 + \|x_v(f)\|^2 - 2\langle x_u(f), x_v(f) \rangle \right)$$

$$= \sum_f \|x_u(f) - x_v(f)\|^2. \qquad \square$$

**Claim 24.**
$$\|X_S^\perp X_u\|^2 \geqslant \sum_f \|P_S^\perp x_u(f)\|^2.$$

*Proof.* For any $\theta \in \mathbb{R}^S$:

$$\|X_u - \sum_v \theta_v X_v\|^2 = \sum_{i=1}^m \left\| \sum_f \langle \overline{x_u(f)}, e_i \rangle x_u(f) - \underbrace{\sum_{v \in S, g} \theta_v \langle \overline{x_v(g)}, e_i \rangle x_v(g)}_{P_S \Theta_i} \right\|^2. \quad (10)$$

Substituting $\alpha_f = P_S^\perp x_u(f)$ and $\beta_f = P_S x_u(f)$, Equation (10) is equal to:

$$= \sum_{i=1}^m \left\| \sum_f \langle \overline{x_u(f)}, e_i \rangle (\alpha_f + \beta_f) - P_S \Theta_i \right\|^2$$

$$= \sum_{i=1}^m \left\| \sum_f \langle \overline{x_u(f)}, e_i \rangle \alpha_f \right\|^2 + \left\| \sum_f \langle \overline{x_u(f)}, e_i \rangle \beta_f - P_S \Theta_i \right\|^2$$

$$\geqslant \sum_{i=1}^m \left\| \sum_f \langle \overline{x_u(f)}, e_i \rangle \alpha_f \right\|^2$$

$$= \sum_{f,g} \langle \alpha_f, \alpha_g \rangle \sum_{i=1}^m \langle \overline{x_u(f)}, e_i \rangle \langle \overline{x_u(g)}, e_i \rangle$$

$$= \sum_{f,g} \langle \alpha_f, \alpha_g \rangle \langle \overline{x_u(f)}, \overline{x_u(g)} \rangle = \sum_f \|\alpha_f\|^2 = \sum_f \|P_S^\perp x_u(f)\|^2. \qquad \square$$

This concludes the proof of Theorem 21, therefore also the proof of Theorem 20.

# 7 The Main Rounding Algorithm and Its Analysis

In this section we state and prove the main results concerning our rounding algorithm for Lasserre SDP solutions, and in particular prove Theorem 31 which we used to analyze our algorithm for quadratic integer programming and its applications to graph partitioning. Some of this discussion already appeared in the simpler setting of Minimum Bisection in Section 3. All our rounding algorithms are based on choosing labels of a carefully chosen "seed" set $\mathcal{S}^*$ of appropriate size $r'$, which is then propagated to other nodes conditioned on the particular labeling of $\mathcal{S}^*$.

For easy reference, we describe the rounding procedure in Algorithm 1 and the seed selection procedure in Algorithm 2.

## 7.1 Simple lemmas about rounding

We first describe how to perform the rounding *after* a good choice of the seed set $\mathcal{S}^*$ has been made, followed by an analysis of its properties. This part is quite simple; the crux of our rounding

**Algorithm 1** Algorithm for labeling in time $O\left(k^{r'} + n\right)$.

**Input:** $\mathcal{S}^* \subseteq V$ of size at most $r'$, $x \in \mathbf{Lasserre}^{(r')}(V \times [k])$.
**Output:** $\widetilde{x} \in \{0,1\}^{V \times [k]}$.
**Procedure:**

1. Choose $f \in [k]^{\mathcal{S}^*}$ with probability $\|x_{\mathcal{S}^*}(f)\|^2$.

2. Label every node $u \in V$ by choosing a label $j \in [k]$ with probability $\frac{\langle x_{\mathcal{S}^*}(f), x_u(j)\rangle}{\|x_{\mathcal{S}^*}(f)\|^2}$.

---

**Algorithm 2** Algorithm for finding seed set in time $O(n^5)$ deterministically.

**Input:** Positive integers $r$, $r' = \frac{r}{\varepsilon^2}$, $x \in \mathbf{Lasserre}^{(r')}(V \times [k])$ and a PSD matrix $L \in \mathbb{R}^{(V\times[k])\times(V\times[k])}$.
**Output:** Seed set $\mathcal{S}^* \subseteq V$ of size at most $r'$ satisfying Equation (11).
**Procedure:**

1. Let $\mathcal{S}^* \leftarrow \emptyset$.

2. Repeat until $\mathcal{S}^*$ satisfies Equation (11):

   (a) Find new $\frac{r}{\varepsilon}$-many seeds $\widetilde{T} \in \binom{V \times [k]}{r/\varepsilon}$ using deterministic column selection algorithm given in [GS11] on matrix $\mathrm{diag}(L)^{1/2}\Pi^\perp_{\mathcal{S}^*}\mathcal{X}$.

   (b) $T \leftarrow \left\{u \,\middle|\, \exists j \in [k] : (u,j) \in \widetilde{T}\right\}$.

   (c) $\mathcal{S}^* \leftarrow \mathcal{S}^* \bigcup T$.

---

is how to choose the best $\mathcal{S}^*$ and bound the performance when it is used as the seed set. This will be described in Section 7.2.

**Definition 25** (Rounding distribution). *Given $x \in \mathbf{Lasserre}^{(r')}(V \times [k])$ and $\mathcal{S}^* \in \binom{V}{r'}$, we define $\mathcal{D}^*$ as the distribution on labellings of $\mathcal{S}^*$, in which a labeling $f \in [k]^{\mathcal{S}^*}$ is chosen with probability:*

$$\Pr_{f'\sim\mathcal{D}^*}\left[f' = f\right] = \|x_{\mathcal{S}^*}(f)\|^2.$$

*Here $f \sim \mathcal{D}^*$ denotes choosing from distribution $\mathcal{D}^*$.*

*For any $f \in [k]^{\mathcal{S}^*}$, we use $\mathcal{D}^*_f$ as the distribution on binary vectors corresponding to labellings of $V$, $\{0,1\}^{V[k]}$, in which each node $u \in V$ receives, independently at random, a label $g \in [k]^u$ with probability:*

$$\Pr_{\widetilde{x}\sim\mathcal{D}^*_f}\left[\widetilde{x}_u(g) = 1\right] = \frac{\|x_{\mathcal{S}^*\cup\{u\}}(f \circ g)\|^2}{\|x_{\mathcal{S}^*}(f)\|^2} = \frac{\langle \overline{x_{\mathcal{S}^*}(f)}, x_u(g)\rangle}{\|x_{\mathcal{S}^*}(f)\|}.$$

*We will abuse the notation and use $\widetilde{x} \sim \mathcal{D}^*$ for sampling a binary labeling vector by first choosing $f \sim \mathcal{D}^*$ and then choosing $\widetilde{x} \sim \mathcal{D}^*_f$.*

We now prove some simple properties of this rounding. All claims below hold for every fixed choice of $\mathcal{S}^*$.

**Claim 26.** *For any $u \in V$ and $g \in [k]^u$, we have*
$$\Pr_{\widetilde{x} \sim \mathcal{D}^*} [\widetilde{x}_u(g) = 1] = \|x_u(g)\|^2.$$

*Proof.* Indeed, by definition of the rounding scheme, $\Pr_{\widetilde{x} \sim \mathcal{D}^*} [\widetilde{x}_u(g) = 1]$ equals
$$\sum_f \|x_{\mathcal{S}^*}(f)\|^2 \frac{\|x_{f \cup \{u\}}(f \circ g)\|^2}{\|x_{\mathcal{S}^*}(f)\|^2} = \sum_f \langle x_{\mathcal{S}^*}(f), x_u(g) \rangle = \langle x_\emptyset, x_u(g) \rangle = \|x_u(g)\|^2 . \quad \square$$

Before stating the next claim, let us again recall the definition of the projection operator used in the analysis of the rounding.

**Definition 27.** *Given $x \in \mathbf{Lasserre}^{(r)}(V \times [k])$, we define $\Pi_S \in \mathbb{R}^{\Upsilon \times \Upsilon}$ as the projection matrix onto the span of $\{x_S(f)\}_{f \in [k]^S}$ for given $S$:*
$$\Pi_S \triangleq \sum_{f \in [k]^S} \overline{x_S(f)} \cdot \overline{x_S(f)}^T.$$

*Define $\Pi_S^\perp = I - \Pi_S$ to be the projection matrix onto the orthogonal complement of the span of $\{x_S(f)\}_{f \in [k]^S}$, where $I$ denotes the identity matrix of appropriate dimension.*

**Claim 28.** *For any $u \neq v \in V$ and $g \in [k]^u, h \in [k]^v$:*
$$\Pr_{\widetilde{x} \sim \mathcal{D}^*} [\widetilde{x}_u(g) = 1 \wedge \widetilde{x}_v(h) = 1] = \langle \Pi_{\mathcal{S}^*} x_u(g), \Pi_{\mathcal{S}^*} x_v(h) \rangle.$$

*Proof.*
$$\Pr_{\widetilde{x} \sim \mathcal{D}^*} [\widetilde{x}_u(g) = 1 \wedge \widetilde{x}_v(h) = 1] = \sum_f \|x_{\mathcal{S}^*}(f)\|^2 \frac{\|x_{\mathcal{S}^* \cup \{u\}}(f \circ g)\|^2 \|x_{\mathcal{S}^* \cup \{v\}}(f \circ h)\|^2}{\|x_{\mathcal{S}^*}(f)\|^4}$$
$$= \sum_f \frac{\|x_{\mathcal{S}^* \cup \{u\}}(f \circ g)\|^2 \|x_{\mathcal{S}^* \cup \{v\}}(f \circ h)\|^2}{\|x_{\mathcal{S}^*}(f)\|^2}$$
$$= \sum_f \frac{\langle x_{\mathcal{S}^*}(f), x_u(g) \rangle \langle x_{\mathcal{S}^*}(f), x_v(h) \rangle}{\|x_{\mathcal{S}^*}(f)\|^2}$$
$$= \sum_f \langle \overline{x_{\mathcal{S}^*}(f)}, x_u(g) \rangle \langle \overline{x_{\mathcal{S}^*}(f)}, x_v(h) \rangle$$
$$= \sum_f x_u(g)^T \overline{x_{\mathcal{S}^*}(f)} \cdot \overline{x_{\mathcal{S}^*}(f)}^T x_v(h)$$
$$= x_u(g)^T \Pi_{\mathcal{S}^*} x_v(h) = \langle \Pi_{\mathcal{S}^*} x_u(g), \Pi_{\mathcal{S}^*} x_v(h) \rangle. \quad \square$$

**Claim 29.** *Given any $\mathcal{S}^*$ with $\widetilde{x}$ sampled from $\mathcal{D}^*$ as described, the following identity holds:*

*For any matrix $L \in \mathbb{R}^{(V \times [k])^2}$,*
$$\mathbb{E}_{\widetilde{x} \sim \mathcal{D}^*} [\widetilde{x}^T L \widetilde{x}] = \mathsf{Tr}(\mathcal{X}^T \Pi_{\mathcal{S}^*}^\perp \mathcal{X} \mathrm{diag}(L)) + \mathsf{Tr}(\mathcal{X}^T \Pi_{\mathcal{S}^*} \mathcal{X} L)$$

*Proof.* Consider $L = \text{diag}(A) + L^o$:

$$\mathbb{E}_{\widetilde{x}\sim\mathcal{D}^*}\left[\widetilde{x}^T L \widetilde{x}\right] = \mathbb{E}_{\widetilde{x}\sim\mathcal{D}^*}\left[\widetilde{x}^T \text{diag}(L)\widetilde{x} + \widetilde{x}^T L^o \widetilde{x}\right]$$

Using Claim 26 and Claim 28:

$$\begin{aligned}
&= \text{Tr}(\mathcal{X}^T\mathcal{X}\text{diag}(L)) + \text{Tr}(\mathcal{X}^T\Pi_{\mathcal{S}^*}\mathcal{X}L^o) \\
&= \text{Tr}(\mathcal{X}^T\mathcal{X}\text{diag}(L)) + \text{Tr}(\mathcal{X}^T\Pi_{\mathcal{S}^*}\mathcal{X}(L - \text{diag}(L))) \\
&= \text{Tr}(\mathcal{X}^T\Pi_{\mathcal{S}^*}^\perp\mathcal{X}\text{diag}(L)) + \text{Tr}(\mathcal{X}^T\Pi_{\mathcal{S}^*}\mathcal{X}L).
\end{aligned}$$
□

## 7.2 Choosing A Good Seed Set

In this section we show how to pick a good $\mathcal{S}^*$ and prove our main result, Theorem 31, which lets us relate the performance of our rounding algorithm to the objective value of relaxation.

We begin with a lemma relating the best bound achieved by column-selection for a matrix $X$ (as in Theorem 7) to the objective function $\text{Tr}(X^T X L)$ with respect to an arbitrary PSD matrix $L$.

**Lemma 30.** *Given $X \in \mathbb{R}^{m \times n}$ and a PSD matrix $L \in \mathbb{R}^{n \times n} \succeq 0$ for any positive integer $r$ and positive constant $\varepsilon > 0$, there exists $r/\varepsilon$ columns, $S \in \binom{[n]}{r}$ of $X$ such that*

$$\text{Tr}(X^T X_S^\perp X \text{diag}(L)) \leqslant \frac{\text{Tr}(X^T X L)}{(1-\varepsilon)\lambda_{r+1}[L; \text{diag}(L)]}$$

*where $\lambda_{r+1}[L; \text{diag}(L)]$ is $(r+1)^{th}$ smallest generalized eigenvalue as defined in Definition 12. Furthermore such $S$ can be found in deterministic $O(rn^4)$ time.*

*Proof.* Let $\widetilde{X} \leftarrow X\text{diag}(L)^{1/2}$ and $\mathcal{L} \leftarrow \text{diag}(L)^{-1/2}L\text{diag}(L)^{-1/2}$ with convention $0/0 = \infty$ and $0 \cdot \infty = 0$. Note that the $i^{th}$ smallest eigenvalue of $\mathcal{L}$, $\lambda_i(\mathcal{L})$, corresponds to the $i^{th}$ smallest generalized eigenvalue $\lambda_i[L; \text{diag}(L)]$ by Observation 13. If we let $\sigma_i$ be $i^{th}$ largest eigenvalue of $\widetilde{X}^T\widetilde{X}$, then using Theorem 7 on vectors $\widetilde{X}$, we can find $S \in \binom{V}{r'}$ in time $O(rn^4)$ such that

$$\text{Tr}(\widetilde{X}^T \widetilde{X}_S^\perp \widetilde{X}) \leqslant \frac{1}{1-\varepsilon} \sum_{i \geqslant r+1} \sigma_i.$$

By the von Neumann-Birkhoff theorem, $\text{Tr}(\widetilde{X}^T \widetilde{X}\mathcal{L})$ is minimized when the $i^{th}$ largest eigenvector of $\widetilde{X}^T\widetilde{X}$ corresponds to the $i^{th}$ smallest eigenvector of $\mathcal{L}$:

$$\text{Tr}(\widetilde{X}^T\widetilde{X}\mathcal{L}) \geqslant \sum_i \sigma_i \lambda_i \geqslant \sum_{i \geqslant r+1} \sigma_i \lambda_i \geqslant \lambda_{r+1} \sum_{i \geqslant r+1} \sigma_i \geqslant (1-\varepsilon)\lambda_{r+1}\text{Tr}(\widetilde{X}^T\widetilde{X}_S^\perp\widetilde{X}).$$

The span of $\{\widetilde{X}_u\}_{u\in S}$ is the same with $\{X_u\}_{u\in S}$ since $\widetilde{X}_u$ differs from $X_u$ only by a scaling factor which does not affect the span. In particular, $\widetilde{X}_S^\perp = X_S^\perp$:

$$\text{Tr}(\widetilde{X}^T \widetilde{X}_S^\perp \widetilde{X}) = \text{Tr}(\widetilde{X}^T X_S^\perp \widetilde{X}) = \text{Tr}(X^T X_S^\perp X \text{diag}(L)).$$

The proof is complete by noting that $\text{Tr}(\widetilde{X}^T\widetilde{X}\mathcal{L}) = \text{Tr}(X^T X L)$. □

**Theorem 31** (Main technical theorem). *Given positive integer $r$ and $\varepsilon \in (0,1)$, let $x$ be a set of vectors satisfying $r' = O\left(\frac{r}{\varepsilon^2}\right)$ rounds of Lasserre hierarchy constraints, $x \in \mathbf{Lasserre}^{(r')}(V \times [k])$ with $\mathcal{X} \triangleq (x_u(f))_{u \in V, f \in [k]^u}$ being the matrix whose columns are vectors of $x$ corresponding to singletons.*

*Given any PSD matrix $L \in \mathbb{R}^{(V \times [k]) \times (V \times [k])}$, with $L \succeq 0$, we can find a seed set $\mathcal{S}^*$ of size at most $r'$ in deterministic time $O(n^5)$ with the following properties. For $\widetilde{x}$ randomly sampled from the distribution $\mathcal{D}^*$, $\widetilde{x} \sim \mathcal{D}^*$, as described in Definition 25:*

1. *$\widetilde{x}$ is a binary vector, $\widetilde{x} \in \{0,1\}^{V \times [k]}$.*

2. *$\widetilde{x}$ is an indicator function of a proper labeling of $V$. In particular for any $u \in V$,*
$$\sum_{i \in [k]} \widetilde{x}_u(i) = 1.$$

3. *If there exists $u$ and $i$ such that $x_u(i) = x_\emptyset$ (equivalently $\|x_u(i)\|^2 = 1$) then $\widetilde{x}_u(i)$ is always $1$. Similarly, if $x_u(i) = 0$, then $\widetilde{x}_u(i)$ is always $0$.*

4. *The expected correlation of $\widetilde{x}$ with $L$ is bounded by the correlation of $x$ with $L$ as follows:*
$$\mathbb{E}_{\widetilde{x} \sim \mathcal{D}^*}\left[\widetilde{x}^T L \widetilde{x}\right] \leqslant \frac{1+\varepsilon}{1-\varepsilon} \frac{\mathsf{Tr}(\mathcal{X}^T \mathcal{X} L)}{\min\{\lambda_{r+1}, 1\}}$$

*Here $\mathcal{X}$ denotes the matrix with columns $x_u(i)$, $(u,i) \in V \times [k]$, and $\lambda_{r+1} = \lambda_{r+1}[L; \operatorname{diag}(L)]$ is $(r+1)^{th}$ smallest generalized eigenvalue as defined in Definition 12.*

*Furthermore this set $\mathcal{S}^*$ satisfies the following bound*
$$\mathsf{Tr}(\mathcal{X}^T \Pi_{\mathcal{S}^*}^\perp \mathcal{X} \operatorname{diag}(L)) + \mathsf{Tr}(\mathcal{X}^T \Pi_{\mathcal{S}^*} \mathcal{X} L) \leqslant \frac{1+\varepsilon}{1-\varepsilon} \frac{\mathsf{Tr}(\mathcal{X}^T \mathcal{X} L)}{\min\{\lambda_{r+1}, 1\}} \quad (11)$$

*where $\Pi_{\mathcal{S}^*}$ is defined as in Definition 27.*

*Proof.* Note that the first three properties follow by construction of $\mathcal{D}^*$. Using Claim 29, it can be seen that the bound (11) is a equivalent to 4. Therefore it suffices to prove item 4.

Let $r_0 \triangleq r/\varepsilon$. Consider picking our "seed" nodes in the following iterative way as described in Algorithm 2. Starting with $S(0) \leftarrow \emptyset$, for each $i \in \{1, 2, \ldots\}$, let
$$\widetilde{S}(i) \leftarrow \underset{S \in \binom{V \times [k]}{r_0}}{\arg\min} \mathsf{Tr}(X(i-1)^T X(i-1)_S^\perp X(i-1) \operatorname{diag}(L)),$$

and let $S(i)$ to be the set of nodes whose at least one label appears in $\widetilde{S}(i)$,
$$S(i) \leftarrow \left\{u \mid \exists f \in [k]^u \text{ such that } (u,f) \in \widetilde{S}(i)\right\}; \quad S_i \leftarrow \bigcup_{j \leqslant i} S(j).$$

followed by $X(i) \leftarrow \Pi_{S_i}^\perp \mathcal{X}$. At each step we set $\mathcal{S}^* \leftarrow S_i$ and repeat this until $\mathbb{E}_{\widetilde{x} \sim \mathcal{D}^*}\left[\widetilde{x}^T L \widetilde{x}\right]$ is at most $\frac{1+\varepsilon}{1-\varepsilon} \frac{\eta}{\min(1, \lambda_{r+1})}$. Here $\eta \triangleq \mathsf{Tr}(\mathcal{X}^T \mathcal{X} L)$. We will show that this procedure will stop for some $i$ with $i \leqslant \lceil \frac{1}{\varepsilon} \rceil$ in Claim 36.

Note that, by Lasserre constraints, all vectors in $\{x_u(f)\}_{u\in S, f\in [k]^u}$ are linear combinations of vectors in $\{x_S(f)\}_{f\in [k]^S}$. Hence for any subset of nodes $T \subseteq V$ of size at most $r'$, $X^\perp_{T\times[k]} \succeq \Pi^\perp_T$.

For any $i$, using $\mathcal{D}^*(i)$ to denote the distribution at iteration $i$ with seed set chosen as $\mathcal{S}^* \leftarrow S_i$, by Claim 29:

$$\mathbb{E}_{\widetilde{x}\sim \mathcal{D}^*(i)}\left[\widetilde{x}^T L \widetilde{x}\right] = \mathsf{Tr}(\mathcal{X}^T \Pi^\perp_{S_i} \mathcal{X}\mathrm{diag}(L)) + \mathsf{Tr}(\mathcal{X}^T \Pi_{S_i} \mathcal{X} L) \tag{12}$$

Let $\xi_i$ be defined as $\xi_i \triangleq \mathbb{E}_{\widetilde{x}\sim\mathcal{D}^*(i)}\left[\widetilde{x}^T L\widetilde{x}\right]$ so that:

$$\xi_i = \underbrace{\mathsf{Tr}(\mathcal{X}^T \Pi^\perp_{S_i} \mathcal{X}\mathrm{diag}(L))}_{\delta_i} + \underbrace{\mathsf{Tr}(\mathcal{X}^T \Pi_{S_i} \mathcal{X}L)}_{\eta_i}$$

Finally for convenience we define $\lambda'_{r+1}$ as the following:

$$\lambda'_{r+1} \triangleq (1-\varepsilon)\min\{\lambda_{r+1}[L;\mathrm{diag}(L)],1\}. \tag{13}$$

We will show that this procedure will stop for some $i$ with $i \leq \lceil \frac{1}{\varepsilon} \rceil$ in Claim 36. Note that each iteration takes time at most $O(r_0 n^4)$. If this procedure takes $K$ iterations, we have $r_0 K \leq n$, hence running time is $O(K r_0 n^4) = O(n^5)$.

**Observation 32.**

$$\delta_{i+1} = \mathsf{Tr}(\mathcal{X}^T \Pi^\perp_{S_{i+1}} \mathcal{X}\mathrm{diag}(L)) \leq \mathsf{Tr}(X(i)^T \Pi^\perp_{S(i+1)} X(i)\mathrm{diag}(L))$$

*Proof.* Note that $\Pi^\perp_{S_i}\Pi^\perp_{S(i+1)}\Pi^\perp_{S_i} \succeq \Pi^\perp_{S_{i+1}}$ since all vectors of the form $x_{S_i}(\cdot)$ and $x_{S(i)}(\cdot)$ are linear combinations of vectors $x_{S_{i+1}}(\cdot)$. Using the definition of $X(i)$, $X(i) = \Pi^\perp_{S_i}\mathcal{X}$, the proof is complete. □

**Observation 33.** *For any $i \geq 0$, we have $\eta_i \leq \eta$.*

*Proof.* Note $\Pi_{S_i} \preceq I$. Since $L$ is PSD, $\eta_i = \mathsf{Tr}(X^T \Pi_{S_i} XL) \leq \mathsf{Tr}(X^T XL) = \eta$. □

**Claim 34.** *For any $i \geq 0$,*

$$\delta_{i+1} \leq \frac{\eta - \eta_i}{\lambda'_{r+1}}.$$

*where $\lambda'_{r+1}$ is defined as in Equation* (13).

*Proof.* Using Observation 32,

$$\begin{aligned}
\delta_{i+1} &\leq \mathsf{Tr}(X(i)^T \Pi^\perp_{S(i+1)} X(i)\mathrm{diag}(L)) \\
&\leq \mathsf{Tr}(X(i)^T X^\perp_{S(i+1)\times[k]} X(i)\mathrm{diag}(L)) \\
&\leq \mathsf{Tr}(X(i)^T X^\perp_{\widetilde{S}(i+1)} X(i)\mathrm{diag}(L)) \\
&\leq \frac{1}{(1-\varepsilon)\lambda_{r+1}}\mathsf{Tr}(X(i)^T X(i)L),
\end{aligned}$$

where the first inequality follows from $\Pi^\perp_{\widetilde{S}(i+1)} \preceq X^\perp_{\widetilde{S}(i+1)\times[k]}$, and the second inequality from $\widetilde{S}(i+1) \subseteq S(i+1) \times [k]$. For the last inequality, we can immediately apply the bound from Lemma 30. Using $(1-\varepsilon)\lambda_{r+1}[L; \mathrm{diag}(L)] \geq \lambda'_{r+1}$ for $\lambda'_{r+1}$ as defined in Equation (13), and the identity

$$\mathsf{Tr}(X(i)^T X(i) L) = \mathsf{Tr}(\mathcal{X}^T \Pi^\perp_{\widetilde{S}_i} \mathcal{X} L) = \mathsf{Tr}(\mathcal{X}^T \mathcal{X} L) - \mathsf{Tr}(\mathcal{X}^T \Pi_{\widetilde{S}_i} \mathcal{X} L) = \eta - \eta_i$$

we conclude the proof. $\square$

**Claim 35.** *If $\xi_{i+1} > \eta \frac{1+\varepsilon}{\lambda'_{r+1}}$, then*

$$\frac{\varepsilon + \eta_i}{\lambda'_{r+1}} < \eta_{i+1}.$$

*Proof.* Using Claim 34,

$$\eta \frac{1+\varepsilon}{\lambda'_{r+1}} < \xi_{i+1} = \delta_{i+1} + \eta_{i+1} \leq \frac{\eta - \eta_i}{\lambda'_{r+1}} + \eta_{i+1}.$$

Hence

$$\frac{\varepsilon + \eta_i}{\lambda'_{r+1}} < \eta_{i+1} \ . \qquad \square$$

**Claim 36.** *There exists $i \leq \lceil \frac{1}{\varepsilon} \rceil$ for which $\xi_i \leq \eta \frac{1+\varepsilon}{\lambda'_{r+1}}$.*

*Proof.* By contradiction. Let $K = \lceil \frac{1}{\varepsilon} \rceil$ and assume for all $i \leq K$, $\xi_i > \eta \frac{1+\varepsilon}{\lambda'_{r+1}}$. By Claim 35,

$$\eta_1 > \eta \frac{\varepsilon}{\lambda'_{r+1}} \geq \eta\varepsilon$$

$$\eta_2 > \eta \frac{\varepsilon}{\lambda'_{r+1}} + \frac{\eta_1}{\lambda'_{r+1}} > \eta \frac{\varepsilon}{\lambda'_{r+1}} (1+1) \geq \eta \cdot 2\varepsilon$$

$$\vdots$$

$$\eta_K > \eta \frac{\varepsilon}{\lambda'_{r+1}} + \frac{\eta_{K-1}}{\lambda'_{r+1}} > \eta K \frac{\varepsilon}{\lambda'_{r+1}} \geq \eta \cdot K\varepsilon \implies \eta_K > \eta.$$

which contradicts Observation 33. $\square$

This completes the proof of Theorem 31. $\square$

# 8 Algorithm for Independent Set

Our final algorithmic result is on finding independent sets in a graph. For simplicity, we focus on unweighted graphs though the extension for graphs with non-negative vertex weights is straightforward. As usual, we denote by $\alpha(G)$ the size of the largest independent set in $G$. Let $d_{\max}$ denote the maximum degree of a vertex of $G$.

**Theorem 37.** *Given $0 < \varepsilon < 1$, positive integer $r$, a graph $G$ with $d_{\max} \geq 3$, there exists an algorithm to find an independent set $I \subseteq V$ such that*

$$|I| \geq \alpha(G) \cdot \min\left\{\frac{1}{2d_{\max}}\left(\frac{1}{(1-\varepsilon)\min\{2-\lambda_{n-r-1}(\mathcal{L}),1\}}-1\right)^{-1}, 1\right\} \quad (14)$$

*in time $n^{O\left(\frac{r}{\varepsilon^2}\right)}$.*

**Remark 10.** The above bound (14) implies that if $\lambda_{n-r-1}$, which is the $(r+1)^{\text{st}}$ largest eigenvalue of the normalized Laplacian $\mathcal{L}$, is very close to 1, then we can find large independent sets in $n^{O(r/\varepsilon^2)}$ time. In particular, if it is at most $1 + \frac{1}{4d_{\max}}$, then taking $\varepsilon = O(1/d_{\max})$, we can find an optimal independent set. The best approximation ratio for independent set in terms of $d_{\max}$ is about $O\left(\frac{d_{\max} \cdot \log \log d_{\max}}{\log d_{\max}}\right)$ [Hal98, Hal02]. The bound (14) gives a better approximation ratio when $\lambda_{n-r-1} \leq 1 + O\left(\frac{1}{\log d_{\max}}\right)$. □

*Proof. (of Theorem 37)* Consider the following integer program for finding largest independent set in $G$:

$$\begin{aligned}
\max \quad & \sum_u \widetilde{x}_u(1) \\
\text{subject to} \quad & \widetilde{x}_u(1)\widetilde{x}_v(1) = 0 \quad \text{for any edge } e = (u,v) \in E, \\
& \widetilde{x}_u(1) + \widetilde{x}_u(2) = 1 \quad \text{for all } u \in V. \\
& \widetilde{x} \in \{0,1\}^{V \times [2]}.
\end{aligned}$$

Solve the corresponding Lasserre relaxation using Theorem 39 with $\varepsilon_0 = \frac{1}{n^{O(1)}}$. Let $x$ be the found solution. Note that every quadratic constraint corresponds to a monomial constraint.

From now on, we assume $x$ is a feasible solution to the following problem

$$\begin{aligned}
\max \quad & \text{Tr}(\mathcal{X}(1)^T \mathcal{X}(1)) \\
\text{subject to} \quad & \langle x_u(1), x_v(1) \rangle = 0 \quad \text{for any edge } e = (u,v) \in E, \\
& x \in \textbf{Lasserre}^{(r')}(V \times [2])
\end{aligned}$$

with value $\sum_u \|x_u(1)\|^2 = \text{Tr}(\mathcal{X}(1)^T \mathcal{X}(1)) = \mu$. Here, $\mathcal{X}(1)$ denotes the matrix with columns $x_u(1)$, and henceforth in the proof we will denote $\mathcal{X}(1)$ by $X$, $X \triangleq \mathcal{X}(1)$.

For $\mathcal{S}^*$ to be chosen later, pick $\widetilde{x} \sim \mathcal{D}^*$ as per Definition 25. For ease of notation, we will denote $\mathcal{S}^*$ by $S$ in the ensuing calculations. We convert $\widetilde{x}$ into an independent set as follows.

1. For each $u$, if $\widetilde{x}_u(1) = 1$ then let $I \leftarrow I \cup \{u\}$ with probability $p_u$ which we will specify later.

2. After the first step, for each edge $e = \{u,v\}$, if $\{u,v\} \subseteq I$, we choose one end point randomly, say $u$, and set $I \leftarrow I \setminus \{u\}$.

By construction, the final set $I$ is an independent set.

Note that for any $u$, the probability that $u$ will be in the final independent set $I$ is at least:

$$\Pr\left[u \in I\right] \geqslant \mathbb{E}\left[p_u \widetilde{x}_u(1)\right] - \frac{1}{2}\mathbb{E}\left[\sum_{v \in N(u)} p_u p_v \widetilde{x}_u(1)\widetilde{x}_v(1)\right]$$
$$= p_u\|x_u(1)\|^2 - \frac{1}{2}\sum_{v \in N(u)} p_u p_v \langle \Pi_S x_u(1), \Pi_S x_v(1)\rangle. \qquad (15)$$

By (15), the expected size of the independent set found by the algorithm satisfies

$$\mathbb{E}\left[|I|\right] \geqslant \sum_u p_u \|x_u(1)\|^2 - \sum_{\{u,v\}\in E} p_u p_v \langle \Pi_S x_u(1), \Pi_S x_v(1)\rangle. \qquad (16)$$

Note that for every edge $\{u,v\} \in E$,

$$\langle \Pi_S x_u(1), \Pi_S x_v(1)\rangle = \sum_{f \in [2]^S} \frac{\langle x_S(f), x_u(1)\rangle \langle x_S(f), x_v(1)\rangle}{\|x_S(f)\|^2} \geqslant 0. \qquad (17)$$

We now consider two cases.

**Case 1:** $\langle \Pi_S x_u(1), \Pi_S x_v(1)\rangle = 0$ for all edges $\{u,v\} \in E$. In this case, we take $p_u = 1$ for all $u \in V$, and by (16), we find an independent set of expected size at least $\mu \geqslant \alpha(G)$.

**Case 2:** In this case, we have

$$\sum_{\{u,v\}\in E} \langle \Pi_S x_u(1), \Pi_S x_v(1)\rangle = \frac{1}{2}\mathsf{Tr}(X^T \Pi_S X A) > 0, \qquad (18)$$

where $A$ is the adjacency matrix of $G$. Let $\mathcal{A} = D^{-1/2}AD^{-1/2}$ be the normalized adjacency matrix, and define

$$\xi \triangleq \frac{\mathsf{Tr}(X^T \Pi_S X \mathcal{A})}{\mathsf{Tr}(X^T X)}. \qquad (19)$$

By (17) and (18), we have $\xi > 0$.

We now pick $p_u = \frac{\alpha}{\sqrt{d_u}}$ for all $u \in V$, where we will optimize the choice of $\alpha$ shortly. For this choice, we have

$$\mathbb{E}\left[|I|\right] \geqslant \alpha \sum_u \frac{1}{\sqrt{d_u}}\|x_u(1)\|^2 - \frac{1}{2}\alpha^2 \mathsf{Tr}(X^T \Pi_S X A)$$
$$\geqslant \frac{\alpha}{\sqrt{d_{\max}}} \sum_u \|x_u(1)\|^2 - \frac{1}{2}\alpha^2 \mathsf{Tr}(X^T \Pi_S X A)$$
$$= \mu\left(\frac{\alpha}{\sqrt{d_{\max}}} - \frac{1}{2}\alpha^2 \underbrace{\frac{\mathsf{Tr}(X^T \Pi_S X \mathcal{A})}{\mathsf{Tr}(X^T X)}}_{\xi}\right)$$

This expression is maximized when $\alpha = \frac{1}{\xi \cdot \sqrt{d_{\max}}}$, for which it becomes:

$$\mathbb{E}\left[|I|\right] \geqslant \frac{\mu}{2d_{\max}}\frac{1}{\xi}. \qquad (20)$$

We now describe how we choose $S = \mathcal{S}^*$. Note that the matrix $I + \mathcal{A}$ is positive semidefinite with diagonal entries equal to 1. By applying Theorem 31 to the matrix $\begin{bmatrix} I + \mathcal{A} & 0 \\ 0 & 0 \end{bmatrix}$, we will choose $S$ such that

$$\begin{aligned}
\mathsf{Tr}(X^T \Pi_S^\perp X) + \mathsf{Tr}(X^T \Pi_S X (I + \mathcal{A})) &\leqslant \frac{\mathsf{Tr}(X^T X (I + \mathcal{A}))}{\lambda'} \\
&= \frac{1}{\lambda'} \mathsf{Tr}(X^T X) \quad \text{since} \quad \mathsf{Tr}(X^T X \mathcal{A}) = 0
\end{aligned}$$

where $\lambda' = (1 - \varepsilon) \min\{\lambda_{r+1}(I + \mathcal{A}), 1\} = (1 - \varepsilon) \min\{2 - \lambda_{n-r-1}(\mathcal{L}), 1\}$.

On the other hand,

$$\begin{aligned}
\frac{\mathsf{Tr}(X^T \Pi_S^\perp X) + \mathsf{Tr}(X^T \Pi_S X (I + \mathcal{A}))}{\mathsf{Tr}(X^T X)} &= \frac{\mathsf{Tr}(X^T \Pi_S^\perp X) + \mathsf{Tr}(X^T \Pi_S X) + \mathsf{Tr}(X^T \Pi_S X \mathcal{A})}{\mathsf{Tr}(X^T X)} \\
&= \frac{\mathsf{Tr}(X^T X) + \mathsf{Tr}(X^T \Pi_S X \mathcal{A})}{\mathsf{Tr}(X^T X)} = 1 + \xi.
\end{aligned}$$

Thus, for such a choice of $S$, we obtain that $\xi \leqslant \frac{1}{\lambda'} - 1$. Substituting this back into Equation (20), we have

$$\mathbb{E}\left[|I|\right] \geqslant \frac{\mu}{2 d_{\max}} \frac{1}{1/\lambda' - 1}. \qquad \square$$

## Acknowledgments

We thank Sivaraman Balakrishnan, Ravishankar Krishnaswamy and Srivatsan Narayanan for useful discussions.

## A  Approximately Solving SDP

In this section, we will show how to solve the SDPs arising from the Lasserre relaxations of problems considered in this paper.

**Notation 38.** *For any matrix $Y \in \mathbb{R}^{M \times M}$, we will use $\vec{Y}$ to denote the "vectorization" of $Y$: Listing each entry of matrix $Y$ linearly in some canonical order. For example, for any two matrix $Y, Z \in \mathbb{R}^{M \times M}$*

$$\mathsf{Tr}(YZ) = \langle \vec{Y}, \vec{Z} \rangle = \vec{Y}^T \vec{Z}.$$

**Theorem 39.** *Consider a QIP minimization problem on $k$-labels with PSD objective matrix, $A \in \mathbb{R}^{(V \times [k]) \times (V \times [k])}$, $A \succeq 0$ of the form*

$$\text{minimize } \widetilde{x}^T A \widetilde{x},$$

*at most $(kn)^{O(1)}$ many linear inequalities $\{(b_i, c_i)\}_i$ of the form*

$$\langle b_i, \widetilde{x} \rangle \geqslant c_i, \text{ with } \|b_i\|_{\max} \leqslant 1,$$

*a set of monomial constraints $\{(T_i, g_i, h_i)\}_i$, where $T_i \subseteq V$, $g_i \in [k]^{T_i}$ and $h_i \in \{0, 1\}$, of the form*

$$\prod_{u \in T_i} \widetilde{x}_u(g_i(u)) = h_i,\ ^{8}$$

*Then for any positive real $\varepsilon_0 > 0$, there exists an algorithm to find a solution $\{x_S(f)\}_{S \in \binom{V}{\leqslant r}, f \in [k]^S}$ for the corresponding $r$-round Lasserre relaxation in time $(kn)^{O(r)} \cdot O(\log(1/\varepsilon_0))$ with the following properties:*

1. *If we let $\mathcal{X}$ denote the matrix with columns corresponding to vectors $x_u(i)$ for all $(u, i) \in V \times [k]$, then*

$$\sum_{u,v,i \in [k], j \in [k]} \langle x_u(i), x_v(j) \rangle A_{(u,i),(v,j)} = \mathsf{Tr}(\mathcal{X}^T \mathcal{X} A) \leqslant \mathsf{Opt} + \varepsilon_0$$

   *where $\mathsf{Opt}$ denotes the optimum value for the SDP formulation.*

2. *It satisfies all Lasserre constraints exactly. In particular $x \in \mathbf{Lasserre}^{(r)}(V \times [k])$.*

3. *For any $S \in \binom{V}{\leqslant r}$ and $f \in [k]^S$, if we let $\mathcal{X}_{(S,f)}$ denote the matrix with columns corresponding to vectors $x_{S \cup \{u\}}(f \circ i^u)$, for all $(u, i) \in V \times [k]$, then*

   (a) *For each linear inequality constraint $(b_i, c_i)$*

   $$\sum_{u, i \in [k]} b_i(u, i) \langle x_S(f), x_u(i) \rangle = \mathsf{Tr}(\mathcal{X}_{(S,f)}^T \mathcal{X}_{(S,f)} \mathrm{diag}(b_i)) \geqslant c_i \|x_S(f)\|^2 - \varepsilon_0.$$

   (b) *For each monomial equality constraint $(T_i, g_i, h_i)$ provided that $|S \cup T_i| \leqslant r$,*

   $$\|x_{S \cup T_i}(f \circ g_i)\|^2 = h_i \|x_S(f)\|^2.$$

---

[8] For QIP problems, these are used to express boundary conditions. For independent set problem, such constraints are necessary to express independence constraints, which are of the form $\widetilde{x}_u(1)\widetilde{x}_v(1) = 0$.

*Proof.* We will try to find $X \in \mathbb{R}^{M \times M}$, representing the Gram matrix associated with the Lasserre vectors. The rows and columns of $X$ are associated with pairs $(S, f)$ where $S \in \binom{V}{\leqslant r}$ and $f \in [k]^S$. For any $S, T \in \binom{V}{\leqslant r}$, $f \in [k]^S$ and $g \in [k]^T$, the corresponding entry of $X$, $X_{(S,f),(T,g)}$ represents the inner product $\langle x_S(f), x_T(g) \rangle$. Note that if we let $A = S \cup T$ and $h = f \circ g$, by Lasserre constraints, if $f_{|S \cap T} = g_{|S \cap T}$ then there exists a unique value $z_{(A,h)}$ such that $X_{(S,f),(T,g)} = z_{(A,h)}$, else $X_{(S,f),(T,g)} = 0$.

Let $\mathcal{Z}$ be the affine space of vectors $z$ with coordinates indexed by $(A, h)$ where $A \in \binom{V}{\leqslant 2r}$ and $h \in [k]^A$ satisfying the constraints:

1. $z_\emptyset = 1$,
2. For any $u \in V$, $z_{(u,k)} = 1 - \sum_{j \leqslant k-1} z_{(u,j)}$,
3. For given $S$ and $f \in [k]^S$, for any $(T_i, g_i, h_i)$, if $h_i = 0$, then $z_{(S \cup T_i, f \circ g_i)} = 0$, else if $h_i = 1$, then $z_{(S \cup T_i, f \circ g_i)} = z_{(S,f)}$.

Let $\mathcal{Y}(z)$ denote the matrix whose entry in row $(S, f)$ and column $(T, g)$ where $S, T \in \binom{V}{\leqslant r}$, $f \in [k]^S$ and $g \in [k]^T$, is given by $z_{(A,h)}$ where $A = S \cup T$ and $h = f \circ g$ if $f_{|S \cap T} = g_{|S \cap T}$, and $0$ otherwise. It is easy to check that if $X = \mathcal{Y}(z)$ is positive semidefinite, then all Lasserre consistency constraints are met by any set of vectors whose Gram matrix equals $X$.

We will also assume objective matrix $A$ is extended to $\mathbb{R}^{M \times M}$ by padding with 0's. Then for such a positive semidefinite matrix $X = \mathcal{Y}(z)$ parametrized by $z \in \mathcal{Z}$:

1. Objective function has the form $\mathsf{Tr}(AX)$.
2. Each Lasserre and monomial constraint is always satisfied by construction.
3. Corresponding to each $S$ and $f \in [k]^S$,

    (a) Each linear inequality constraint $\langle b_i, \widetilde{x} \rangle \geqslant c_i$ takes the form $\mathsf{Tr}(B_i X) \geqslant 0$.

From now on we will use $\mathcal{B}$ to denote the matrix whose rows correspond to linear inequality constraints 3a over all $S$ and $f$. Thus the SDP we need to solve can be represented as

$$\begin{aligned} \text{Find } z \text{ minimizing} \quad & \mathsf{Tr}(AX), \\ \text{subject to} \quad & \mathcal{B}\vec{X} \geqslant 0, \\ & X = \mathcal{Y}(z) \succeq 0. \end{aligned}$$

We can use the interior point method [NN06] to solve this problem within an accuracy of $\varepsilon' \triangleq \varepsilon_0 \|B\|_{\min}$ in time $(kn)^{O(r)} \cdot O(\log(1/\varepsilon_0))$ and find $z$ with following properties: For $X = \mathcal{Y}(z)$,

- $\mathsf{Tr}(AX) \leqslant \mathsf{Opt} + \varepsilon_0$,
- $X \succeq 0$,
- $\mathcal{B}\vec{X} \geqslant -\varepsilon_0$.
- (by construction) $X$ satisfies all monomial and Lasserre consistency constraints exactly.

$\square$